\documentclass[twocolumn,english,aps,prl,letterpaper,showpacs,floatfix,groupedaddress,superscriptaddress,utf8]{revtex4-1}

\usepackage[T1]{fontenc}
\usepackage[latin9]{inputenc}
\usepackage{color}
\usepackage{babel}
\usepackage{amsmath}
\usepackage{amssymb}
\usepackage{graphicx}
\usepackage{esint}
\usepackage{subscript}
\usepackage[unicode=true,pdfusetitle,
 bookmarks=true,bookmarksnumbered=false,bookmarksopen=false,
 breaklinks=false,pdfborder={0 0 1},backref=false,colorlinks=true]
 {hyperref}
\usepackage{breakurl}

\makeatletter

\providecommand{\tabularnewline}{\\}

 \@ifundefined{textcolor}{}
 {%
   \definecolor{BLACK}{gray}{0}
   \definecolor{WHITE}{gray}{1}
   \definecolor{RED}{rgb}{1,0,0}
   \definecolor{GREEN}{rgb}{0,1,0}
   \definecolor{BLUE}{rgb}{0,0,1}
   \definecolor{CYAN}{cmyk}{1,0,0,0}
   \definecolor{MAGENTA}{cmyk}{0,1,0,0}
   \definecolor{YELLOW}{cmyk}{0,0,1,0}
 }

\def\part#1{\left(#1\right)}

\usepackage{amsthm}
\usepackage{graphics}
\usepackage{times}
\usepackage{bm}
\usepackage{hyperref}
\usepackage{color}

\definecolor{RoyalBlue}{cmyk}{1, 0.80, 0, 0}

\hypersetup{
colorlinks=true,
breaklinks=true, 
hyperfootnotes=true,
urlcolor=RoyalBlue, 
citecolor=RoyalBlue, %Black
linkcolor=RoyalBlue, %Black 
}

\makeatother

\begin{document}

\title{A Semiclassical ``Divide-and-Conquer'' Method for Spectroscopic
Calculations of High Dimensional Molecular Systems}

\author{Michele \surname{Ceotto}}

\affiliation{Dipartimento di Chimica, Università degli Studi di Milano, via C.
Golgi 19, 20133 Milano, Italy}
\email{michele.ceotto@unimi.it}

\author{Giovanni \surname{Di Liberto}}

\affiliation{Dipartimento di Chimica, Università degli Studi di Milano, via C.
Golgi 19, 20133 Milano, Italy}

\author{Riccardo \surname{Conte}}

\affiliation{Dipartimento di Chimica, Università degli Studi di Milano, via C.
Golgi 19, 20133 Milano, Italy}
\begin{abstract}
A new semiclassical ``divide-and-conquer'' approach is presented
with the aim to demonstrate that quantum dynamics simulations of high
dimensional molecular systems are doable. The method is first tested
by calculating the quantum vibrational power spectra of water, methane
and benzene, three molecules of increasing dimensionality for which
benchmark quantum results are available, and then applied to C\textsubscript{60},
a system characterized by 174 vibrational degrees of freedom. Results
show that the approach can accurately account for quantum anharmonicities,
purely quantum features like overtones, and removal of degeneracy
when the molecular symmetry is broken.
\end{abstract}
\maketitle
\begin{figure}
\begin{centering}
\includegraphics{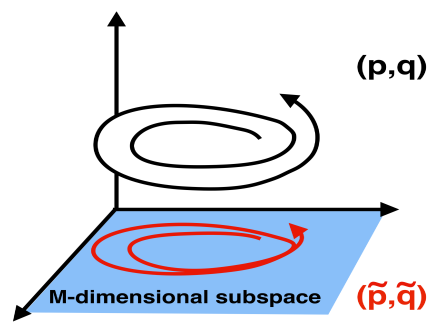}
\par\end{centering}
\caption{\label{fig:idea}Pictorial representation of the projection procedure.}
\end{figure}

Quantum computational approaches to the spectroscopy of small or medium-size
molecules are very popular. Among them we recall variational methods
like vibrational configuration interaction (VCI)\citep{Bowman_Huang_MULTIMODE_2003,Samanta_Reisler_Mixedclusters_2016,Chen_Bowman_Methane-Water_2015,Avila_Carrington_C2H4_2011}
and Multi Configuration Time Dependent Hartree (MCTDH),\citep{manthe2002reaction,Meyer_Worth_HighDimMCTDH_2003,Bowman_Meyer_Polyatomic_2008}
or perturbative ones, such as the second-order vibrational perturbation
theory (VPT2).\citep{Bludsky_Hobza_Anharmonicgly_2000,Barone_AutomatedVPT2_2005,Biczysko_Barone_abinitioIRgly_2012}
Spectroscopy of high dimensional systems is more difficult to perform,
since exact quantum simulations are unaffordable and even experimental
spectra are often too crowded for an undisputed assignment. A new
computationally-affordable strategy is needed, while spectra would
certainly be much easier to read if they were decomposed into several
partial ones.\\\indent  For this purpose, a novel theoretical approach
is here presented. It is based on a semiclassical (SC) ``divide-and-conquer''
strategy that leads to reliable calculations of higher dimensional
systems than those ordinarily affordable with quantum methods. Full
spectra are regained as a collection of partial ones, quantum effects
are included, and a sound spectroscopic interpretation is obtained.
This new method fills in the gap between a purely classical spectroscopic
study, which is not satisfactory because it neglects key quantum features,
and quantum approaches, which often require the set-up of a grid of
points with a computational cost that exponentially scales with the
dimensionality of the system.\\\indent  In a semiclassical approach\citep{Elran_Kay_ImprovingHK_1999,Zhang_Pollak_Deeptunneling_2004,Kay_SCcorrections_2006,Conte_Pollak_ThawedGaussian_2010,Bonella_Kapral_quantum-classical_2010,Monteferrante_Ciccotti_Liquidneon_2013,Conte_Pollak_ContinuumLimit_2012,Petersen_Pollak_Interactionrepresentation_2015,Shalashilin_Child_Coherentstates_2001,Shalashilin_Child_CCS_2004,Heatwole_Prezhdo_Morsesemiclassical_2009,Garashchuk_Prezhdo_Bohmian_2011,Huo_Coker_Semiclassicalnonadiabatic_2012,Harinder_Tomsovic_GeneralizedGaussian_2016,Koch_Tannor_WPrevival_2017,Harabati_Grossmann_LongtimeSCIVR_2004,Grossmann_HierarchySC_1999,Nakamura_Ohta_SCDevelopment_2016,Kondorskiy_Nanbu_Nonadiabatic_2015,Tao_ImportanceSampling_2014,Antipov_Nandini_Mixedqcl_2015,Liu_Miller_ThermalGaussian_2006,Liu_Miller_linearizedSCIVR_2007,Liu_Miller_RealTimecorrelation_2007,Liu_Miller_VariouslinearizedSCIVR_2008,Koda_SCIVRWigner_2015,Koda_Mixedsemiclassical_2016,Ushiyama_Takatsuka_Interferences_2005,Takahashi_Takatsuka_PhaseQuantization_2007,Zhuang_Ceotto_Hessianapprox_2012,Wehrle_Vanicek_Oligothiophenes_2014,Wehrle_Vanicek_NH3_2015,Zambrano_Vanicek_Cellulardephasing_2013}
spectra are calculated in a time-dependent way from classically evolved
trajectories, and, if convenient, pre-computation of the potential\citep{Braams_Bowman_PermutInvariant_2009,Conte_Bowman_GaussianBinning_2013,Jiang_Guo_NeuralNetworks_2014,Conte_Bowman_Communication_2014,Houston_Bowman_CollisionModel_2014,Conte_Bowman_CollisionsCH4-H2O_2015,Conte_Bowman_Manybody_2015,Houston_Bowman_ModelFinal_2015,Houston_Bowman_RoamingH2CO_2016}
can be avoided in favor of a direct dynamics,\citep{Ceotto_AspuruGuzik_Multiplecoherent_2009,Tatchen_Pollak_Onthefly_2009,Ceotto_AspuruGuzik_Curseofdimensionality_2011,Wong_Roy_Formaldehyde_2011,Ceotto_Hase_AcceleratedSC_2013,Wehrle_Vanicek_NH3_2015}
thus allowing to explore the global potential energy surface also
when dealing with high-dimensional systems. Recently, we have advanced
Miller's pivotal semiclassical initial value representation (SCIVR)\citep{Heller_FrozenGaussian_1981,Herman_Kluk_SCnonspreading_1984,Miller_PNAScomplexsystems_2005,Kay_Atomsandmolecules_2005,Miller_Atom-Diatom_1970,Miller_Molecularcollisions_1974}
theory by developing the multiple-coherent (MC) SCIVR approach.\citep{Ceotto_AspuruGuzik_Multiplecoherent_2009,Ceotto_Tantardini_Copper100_2010}
The method exploits pioneering work by De Leon and Heller, which demonstrated
that even single-trajectory semiclassical simulations are able to
precisely reproduce quantum eigenvalues and eigenfunctions.\citep{DeLeon_Heller_SCeigenfunctions_1983}
MC-SCIVR is based on a tailored coherent state semiclassical representation
and yields highly accurate results in spectroscopy calculations, often
within 1\% of the exact result, given a few classical trajectories
as input. Applications have faithfully reproduced a variety of quantum
effects, including quantum resonances, intra-molecular and long-range
dipole splitting, and the quantum resonant umbrella inversion in ammonia.\citep{Ceotto_AspuruGuzik_PCCPFirstprinciples_2009,Ceotto_AspuruGuzik_Firstprinciples_2011,Conte_Ceotto_NH3_2013,Tamascelli_Ceotto_GPU_2014,Buchholz_Ceotto_MixedSC_2016,DiLiberto_Ceotto_Prefactors_2016}
However, the approach runs out of steam when the dimensionality increases
and it is limited to about 20-25 degrees of freedom. \\ \indent To
understand the reasons of such a limitation, we observe that a N-dimensional
semiclassical wavepacket is built as the direct product of monodimensional
coherent states $\left|\chi\left(t\right)\right\rangle =\left|\chi_{1}\left(t\right)\right\rangle ...\left|\chi_{N}\left(t\right)\right\rangle $
and power spectra are obtained as Fourier transforms of the recurring
time-dependent overlap $\left\langle \chi\left(0\right)|\chi\left(t\right)\right\rangle $.
Consequently, for a precise spectral density it is essential that
the time-evolved semiclassical wavepacket significantly overlaps with
its initial guess. More specifically, the multidimensional classical
trajectory must visit phase space configurations $\left(\mathbf{p}_{t},\mathbf{q}_{t}\right)$
that are close enough to the starting one $\left(\mathbf{p}_{0},\mathbf{q}_{0}\right)$.
The curse of dimensionality occurs because \textit{all} the monodimensional
coherent state overlaps ($\left\langle \chi_{i}\left(0\right)|\chi_{i}\left(t\right)\right\rangle $)
should be sizable almost simultaneously, but for oscillators with
non-commensurable frequencies (even if uncoupled) the concomitant
overlapping event is more and more unlikely as the dimensionality
increases. It is here evident the difference between a semiclassical
and a classical simulation based on a dipole-dipole correlation function.
In fact, the dipole is always a three-dimensional vector, so it is
easier to have a substantial time-dependent overlap. \\ \indent Figure
\ref{fig:idea} illustrates how we think to overcome the curse of
dimensionality in semiclassical calculations.
 In few words, a full dimensional classical trajectory (black line)
has higher odds to get close to its initial configuration if projected
onto a subspace (red line). Based on this observation, we propose
that while classical trajectories be still treated in full dimensionality,
the semiclassical calculation employ sub-space bounded information
to yield projected spectra. From a statistical point of view, the
procedure corresponds to the calculation of a marginal distribution
in each subspace after marginalizing out the other degrees of freedom.\citep{Trumpler_Weaver_StatAstronomy_1962}
As a final step, the composition of the several projected spectra
provides the full-dimensional one. 

We apply this idea to spectral density, $I(E)$, calculations
\begin{equation}
I\left(E\right)\equiv\frac{1}{2\pi\hbar}\int_{-\infty}^{+\infty}\left\langle \chi\left|e^{-i\hat{H}t/\hbar}\right|\chi\right\rangle e^{iEt/\hbar}dt.\label{eq: Power_Spectrum}
\end{equation}

\noindent An exact representation of the quantum propagator $e^{-i\hat{H}t/\hbar}$
is given by Feynman's path integral formulation, which can be approximated
by considering only the classical paths connecting points ${\bf q}{}_{0}$
and ${\bf q}^{\prime}$ in time t (roots) and including fluctuations
up to the second order around the classical action ($S^{cl}$) of
each path\citep{Gutzwiller_SCpropagator_1967,VanVleck_SCpropagator_1928}
\begin{equation}
\left\langle \mathbf{q}^{\prime}\left|e^{-\frac{i}{\hbar}\hat{H}t}\right|\mathbf{q}_{0}\right\rangle \approx\sum_{roots}\left[\dfrac{\left|-\frac{\partial^{2}S_{t}^{cl}}{\partial\mathbf{q}^{\prime}\partial\mathbf{q}_{0}}\right|}{\left(2\pi i\hbar\right)^{N}}\right]^{1/2}\dfrac{e^{\frac{i}{\hbar}S_{t}^{cl}\left(\mathbf{q}^{\prime},\mathbf{q}_{0}\right)}}{e^{i\upsilon\pi/2}}.\label{eq:van_Vleck}
\end{equation}
 Eq. (\ref{eq:van_Vleck}) represents the semiclassical approximation
to the Feynman path integral.\citep{Berry_Mount_Semiclassical_1972}
The term $e^{-i\upsilon\pi/2}$, where $\upsilon$ is the integer
Maslov index, ensures the continuity of the square root of the pre-exponential
factor. However, the drawback of Eq. (\ref{eq:van_Vleck}) is the
presence of points at which the determinant in the pre-exponential
factor becomes singular. Miller's SCIVR\citep{Miller_S-Matrix_1970,Heller_CellularDynamics_1991}
overcomes this issue by replacing the sum over classical trajectories
with an integration over initial momenta, a very powerful approach
especially when combined with Heller's coherent states ($|\mathbf{p},\mathbf{q\rangle}$)
representation. Coherent states have a Gaussian coordinate-space representation
whose width is given by the (usually diagonal) $\Gamma$ width matrix
\begin{equation}
\left\langle \mathbf{x}|\mathbf{p},\mathbf{q}\right\rangle =\left(\frac{det\left(\Gamma\right)}{\pi^{N}}\right)^{1/4}e^{-\left(\mathbf{x}-\mathbf{q}\right)^{T}\frac{\Gamma}{2}\left(\mathbf{x}-\mathbf{q}\right)+i\mathbf{p}\left(\mathbf{x}-\mathbf{q}\right)/\hbar}.\label{eq:coherent_state}
\end{equation}
By using Miller's SCIVR and by either reformulating the Feynman paths\citep{Weissman_coherentstates_1982,Baranger_Schellhaass_2001}
or representing the spectral density $I(E)$\citep{Herman_Kluk_SCnonspreading_1984}
in terms of the coherent states of Eq. (\ref{eq:coherent_state}),
one gets to the working formula

\small
\begin{equation}
\begin{aligned}I\left(E\right)=\frac{1}{2\pi\hbar}\int_{-\infty}^{+\infty}dte^{iEt/\hbar} & \frac{1}{\left(2\pi\hbar\right)^{N}}\int\int d\mathbf{q}_{0}d\mathbf{p}_{0}C_{t}\left(\mathbf{p}_{0},\mathbf{q}_{0}\right)\\
 & e^{iS_{t}\left(\mathbf{p}_{0},\mathbf{q}_{0}\right)/\hbar}\left\langle \chi|\mathbf{p}_{t},\mathbf{q}_{t}\right\rangle \left\langle \mathbf{p}_{0},\mathbf{q}_{0}|\chi\right\rangle .
\end{aligned}
\label{eq:Power_Spectrum_HK}
\end{equation}

\normalsize \noindent where

\begin{equation}
C_{t}\left(\mathbf{p}_{0},\mathbf{q}_{0}\right)=\sqrt{\frac{1}{2}\left|\frac{\partial\mathbf{q}_{t}}{\partial\mathbf{q}_{0}}+\frac{\partial\mathbf{p}_{t}}{\partial\mathbf{p}_{0}}-i\hbar\Gamma\frac{\partial\mathbf{q}_{t}}{\partial\mathbf{p}_{0}}+\frac{i}{\Gamma\hbar}\frac{\partial\mathbf{p}_{t}}{\partial\mathbf{q}_{0}}\right|}.\label{eq:prefactor}
\end{equation}

In order to accelerate the Monte Carlo integration of Eq. (\ref{eq:Power_Spectrum_HK}),
it is possible to insert a time averaging filter $\frac{1}{T}\int_{0}^{T}dt$
without loss of accuracy by virtue of Liouville's theorem. Miller
et al.\citep{Kaledin_Miller_TAmolecules_2003,Kaledin_Miller_Timeaveraging_2003}
worked out the following time averaged (TA) version of Eq. (\ref{eq:Power_Spectrum_HK})

\begin{eqnarray}
I\left(E\right) & = & \left(\frac{1}{2\pi\hbar}\right)^{N}\iintop d\mathbf{p}_{0}d\mathbf{q}_{0}\frac{1}{2\pi\hbar T}\nonumber \\
 & \times & \left|\intop_{0}^{T}dte^{\frac{i}{\hbar}\left[S_{t}\left(\mathbf{p}_{0},\mathbf{q}_{0}\right)+Et+\phi_{t}\right]}\left\langle \boldsymbol{\chi}\left|\mathbf{p}_{t}\mathbf{q}_{t}\right.\right\rangle \right|^{2}\label{eq:separable}
\end{eqnarray}

\noindent where the additional approximation $\phi\left(t\right)=\mbox{phase}\left[C_{t}\left(\mathbf{p}_{0},\mathbf{q}_{0}\right)\right]$
has been introduced. Eq. (\ref{eq:separable}) is now much easier
to converge due to its positive-definite integrand, and it has been
tested on several molecules\citep{Kaledin_Miller_TAmolecules_2003,Conte_Ceotto_NH3_2013,Ceotto_AspuruGuzik_Firstprinciples_2011,Gabas_Ceotto_Glycine_2016,Ceotto_AspuruGuzik_Curseofdimensionality_2011}
yielding very accurate results upon evolution of just about $1000$
trajectories per degree of freedom. The interested reader can find
detailed derivations of the above formulae in Ref. \citenum{Tannor_book_2007}
(Chapter 10) or in Ref. \citenum{Suppl_Info_PRL}.

To further reduce the computational overhead to just a handful of
trajectories we have recently developed an implementation of Eq. (\ref{eq:separable})
based on two observations. First, accurate eigenvalues can be extracted
from a single trajectory whose energy not necessarily must be equal
to the exact (but unknown) eigenvalue.\citep{DeLeon_Heller_SCeigenfunctions_1983}
Second, for each spectroscopic peak the most contributing trajectories
are those that evolve in the proximity of the vibrational peak energy
shell.\citep{Ceotto_AspuruGuzik_Curseofdimensionality_2011}\textcolor{black}{{}
Based on these considerations, we employ a reference state }$\left|\chi\right\rangle =\sum_{i=1}^{N_{\text{states}}}\left|\mathbf{p}_{\text{eq}}^{i},\mathbf{q}_{\text{eq}}^{i}\right\rangle $
written as a combination of coherent states placed at the classical
phase space points $\left(\mathbf{p}_{\text{eq}}^{i},\mathbf{q}_{\text{eq}}^{i}\right)$.
$\mathbf{q}_{\text{eq}}^{i}$ indicates the equilibrium configuration
and $\mathbf{p}_{\text{eq}}^{i}$ the corresponding multidimensional
momentum. We set $V\left(\mathbf{q}_{\text{eq}}^{i}\right)=0$, and
$\mathbf{p}_{\text{eq}}^{i}$ is chosen to be made of harmonically
estimated momenta, i.e. $\left(p_{j,\text{eq}}^{i}\right)^{2}/2m=\hbar\omega_{j}\left(n_{j}^{i}+1/2\right)$
for the generic j-th vibrational mode. The set of $\omega_{j}$ is
obtained by diagonalizing the Hessian at the equilibrium configuration.
In this way, we can approximate Eq. (\ref{eq:separable}) to

\begin{equation}
\begin{aligned} & I\left(E\right)=\frac{1}{\left(2\pi\hbar\right)^{N}}\frac{Re}{\pi\hbar T}\sum_{i=1}^{n_{states}}\\
 & \left|\int_{0}^{T}dt\left\langle \sum_{i=1}^{n_{states}}\mathbf{p^{i}}_{eq},\mathbf{q^{i}}_{eq}|\mathbf{p}_{t},\mathbf{q}_{t}\right\rangle e^{i\left(S_{t}\left(\mathbf{p}_{eq}^{i},\mathbf{q}_{eq}^{i}\right)+Et+\phi\left(t\right)\right)/\hbar}\right|^{2}
\end{aligned}
\label{eq:MC-TA-SCIVR_final}
\end{equation}

\noindent where $n_{states}$ classical trajectories are evolved
from the initial conditions $\left(\mathbf{p}_{\text{eq}}^{i},\mathbf{q}_{\text{eq}}^{i}\right)$.
This approach is called Multiple Coherent TA-SCIVR (MC-SCIVR) (also
indicated as MC-TA-SCIVR). MC-SCIVR has been shown to be accurate
for systems of complexity up to the glycine molecule (i.e. 24 degrees
of freedom).\citep{Gabas_Ceotto_Glycine_2016}

The main theoretical novelty presented in this Letter is that we re-formulate
Eq.(\ref{eq:separable}) on the basis of projected-trajectory information.
First, the $N$ dimensional phase space is conveniently partitioned,
i.e. $\left(\mathbf{p},\mathbf{q}\right)\equiv\left(p_{1},q_{1,}...,\tilde{p}_{i},\tilde{q}_{i},...,\tilde{p}_{i+M},\tilde{q}_{i+M},...,p_{N},q_{N}\right)$,
where we have highlighted a generic M-dimensional subspace ($M<N)$
with tilde variables $\left(\tilde{\mathbf{p}},\tilde{\mathbf{q}}\right)$
(see Fig.(\ref{fig:idea})). For this purpose, an analysis is performed
concerning the off-diagonal values of the Hessian matrix averaged
over a full-dimensional classical trajectory with harmonic zero-point
energy. Off-diagonal terms that are bigger than a threshold value
($\varepsilon$) correspond to coupled modes and are included in the
same subspace. The threshold choice is driven by the trade-off between
calculation accuracy and feasibility. On one hand, the smaller the
threshold value the smaller the number of neglected interactions and
the more accurate the calculation. On the other, the dimensionality
of any projected space should not exceed 20-25 degrees of freedom
to permit MC-SCIVR calculations in that subspace. Then, we consider
that each vector or matrix appearing in Eq.(\ref{eq:separable}) can
be exactly projected into each sub-space by means of a singular value
decomposition procedure $\mathbf{A}=\mathbf{U}\mathbf{\Sigma}\mathbf{V}$,\citep{Hinsen_Kneller_SingValueDecomp_2000}
and consequently restrict the phase space integration to $\int\int d\mathbf{\tilde{p}}_{0}d\tilde{\mathbf{q}}_{0}$.
The M-dimensional coherent state becomes
\begin{equation}
\left\langle \mathbf{\tilde{x}}\left|\mathbf{\tilde{p}}_{t}\mathbf{\tilde{q}}_{t}\right.\right\rangle =\left(\frac{\mbox{det}(\tilde{\Gamma})}{\pi^{M}}\right)^{\frac{1}{4}}e^{-\frac{1}{2}\left(\mathbf{\tilde{x}}-\mathbf{\tilde{q}}_{t}\right)^{T}\mathbf{\tilde{\Gamma}}\left(\mathbf{\tilde{x}}-\mathbf{\tilde{q}}_{t}\right)+\frac{i}{\hbar}\mathbf{\tilde{p}}_{t}^{T}\left(\mathbf{\tilde{x}}-\mathbf{\tilde{q}}_{t}\right),}\label{eq:projected_coherent}
\end{equation}
where $\widetilde{\mathbf{\Gamma}}=\mathbf{U}\mathbf{U}^{T}\mathbf{\Gamma}\mathbf{U}\mathbf{U}^{T}$
is the projected Gaussian width matrix obtained from the singular-value
decomposition matrix ${\bf U}$.\citep{Harland_Roy_SCIVRconstrained_2003}
Similarly, $\widetilde{C}_{t}$ is obtained by projecting its monodromy
matrix components. The remaining term of Eq.(\ref{eq:separable})
to be projected is $S_{t}.$ While the projection of the kinetic part
of the Lagrangian can be obtained exactly, the potential is generally
not separable. In an ideal case, $V_{S}\left(\tilde{\mathbf{q}}_{M}\right)$
would be the potential such that, given the initial conditions $\left(\tilde{\mathbf{p}}_{0},\tilde{\mathbf{q}}_{0}\right)$,
the M-dimensional trajectory coincides with the projected one. In
such a M-dimensional dynamics, the positions in the other degrees
of freedom ($\mathbf{q}_{N-M}$) are downgraded to parameters. In
practice, we fix these parameters at equilibrium positions, but introduce
an external field $\lambda\left(t\right)$ to account for the non-separability
of the potential such that 
\begin{equation}
V_{S}\left(\tilde{\mathbf{q}}_{M}\right)\equiv V\left(\tilde{\mathbf{q}}_{M};\mathbf{q}_{N_{vib}-M}\right)=V\left(\tilde{\mathbf{q}}_{M};\mathbf{q}_{N_{vib}-M}^{eq}\right)+\lambda\left(t\right).\label{eq:external_field}
\end{equation}
$\lambda\left(t\right)$ is not known a priori and we adopt the following
expression, which makes Eq. (\ref{eq:external_field}) exact (within
a constant) in the separable potential limit
\begin{eqnarray}
\lambda\left(t\right) & = & V\left(\tilde{\mathbf{q}}_{M};\mathbf{q}_{N_{vib}-M}\right)-\nonumber \\
 &  & \left[V\left(\tilde{\mathbf{q}}_{M};\mathbf{q}_{N_{vib}-M}^{eq}\right)+V\left(\tilde{\mathbf{q}}_{M}^{eq};\mathbf{q}_{N_{vib}-M}\right)\right].\label{eq:lambda}
\end{eqnarray}

Moving to applications, we have first tested accuracy and effectiveness
of our new Divide-and-Conquer Semiclassical Initial Value Representation
(DC-SCIVR) approach on three different molecular systems for which
exact vibrational eigenenergies are available in the literature. 

Water is a low dimensional but strongly coupled system. Its global
3-dimensional vibrational space can be divided into a monodimensional
one for the bending mode, plus a bidimensional one for the two stretches.
We evolved 3500 classical trajectories on a pre-existing potential
energy surface,\citep{Bowman_Zuniga_H2Opotential_1988} each one for
a total of 30000 atomic time units. The zero point energy (ZPE) estimated
from the projected spectra is 4606 $cm^{-1}$, to be compared to the
4631 $cm^{-1}$ value of a full dimensional semiclassical calculation,
and the exact quantum value of 4636 $cm^{-1}$. DC-SCIVR reproduces
fundamentals concerning the bending and the asymmetric stretch with
excellent accuracy (within 10 cm\textsuperscript{-1} of exact quantum
results), while the symmetric stretch and the first bending overtone
are more off the mark (40 cm\textsuperscript{-1}). Overall, the mean
absolute error (MAE) is 23 $cm^{-1}$. Detailed comparisons can be
found in the Supplemental Material.\citep{Suppl_Info_PRL} Results
for water are a remarkable milestone because of the strong internal
vibrational coupling of this molecule. In fact, in higher dimensional
systems inter-mode couplings are generally weaker and DC-SCIVR (being
exact for separable systems) is expected to perform better once strongly
coupled modes are confined into the same subspace.

\begin{table}
\centering{}\caption{Vibrational frequencies of CH\protect\textsubscript{4}. ``QM''
labels the exact quantum eigenvalues; ``SCIVR'' refers to a full
dimensional semiclassical calculation; ``DC-SCIVR'' labels frequencies
obtained with the ``divide-and-conquer'' approach here presented;
``HO'' are harmonic estimates. All values are in $cm^{-1}.$ \label{Table_methane}}
\begin{tabular}{|ccccc|}
\hline 
State & \,QM\textsuperscript{\citep{Carter_Bowman_Methane_1999}} & SCIVR & DC-SCIVR & HO\tabularnewline
\hline 
$1_{1}$ & 1313 & 1300 & 1300 & 1345\tabularnewline
$2_{1}$ & 1535 & 1529 & 1532 & 1570\tabularnewline
$1_{2}$ & 2624 & 2594 & 2606 & 2690\tabularnewline
$1_{1}2_{1}$ & 2836 & 2825 & 2834 & 2915\tabularnewline
$3_{1}$ & 2949 & 2948 & 2964 & 3036\tabularnewline
$2_{2}$ & 3067 & 3048 & 3050 & 3140\tabularnewline
$4_{1}$ & 3053 & 3048 & 3044 & 3157\tabularnewline
MAE &  & 12 & 11 & 68\tabularnewline
\hline 
\end{tabular}
\end{table}

Another known issue for SC methods comes from chaotic trajectories
which can spoil the SC simulation and are therefore usually discarded.
In an application of DC-SCIVR to methane, the 9-dimensional vibrational
space has been partitioned into a 6-dimensional and a 3-dimensional
one and it turns out that methane dynamics is highly chaotic with
strong quantum effects, given the light mass of the hydrogen atoms.
In fact, 95\% of the 180000 trajectories run (each one evolved for
30000 atomic time units) has been discarded on the basis of the monodromy
determinant conservation criterion.\citep{Kaledin_Miller_Timeaveraging_2003,Zhuang_Ceotto_Hessianapprox_2012}
Table \ref{Table_methane} provides a comparison between our DC-SCIVR
estimates and exact values by Bowman on the same analytical surface.\citep{Carter_Bowman_Methane_1999}
This test permits to show that DC-SCIVR works pretty well, with fundamentals
and overtones reliably detected and a tiny MAE (11 cm\textsuperscript{-1}).

Recently, by employing a pre-existing potential energy surface,\citep{Maslen_Jayatilaka_Anharmonicbenzene_1992}
Halverson and Poirier have calculated a set of quantum vibrational
frequencies of benzene with their exact quantum dynamics (EQD) method,\citep{Halverson_Poirier_Benzene_2015}
which we use to benchmark our DC-SCIVR results for this high dimensional
molecular system. For this purpose, the vibrational space of benzene
has been divided into a larger 8-dimensional subspace plus 8 bidimensional
and 6 monodimensional ones. We have evolved 1,000 trajectories per
degree of freedom for a total of 30000 atomic time units each. Furthermore,
an accurate second-order perturbative approximation to the pre-exponential
factor $\exp(i\phi(t)/\hbar)$, as described in Ref. \citenum{DiLiberto_Ceotto_Prefactors_2016},
has been employed to avoid discard of chaotic trajectories. Results
are reported in Table \ref{tab:benzene_results} and permit to assess
DC-SCIVR accuracy in this challenging application. Even in the case
of benzene DC-SCIVR is characterized by a small MAE value (19 cm\textsuperscript{-1}).
This is the result of a large majority of highly accurate frequencies
and a single mode with lower precision. 
\begin{table}
\begin{centering}
\footnotesize
\par\end{centering}
\begin{centering}
\caption{Comparison between DC-SCIVR and available quantum results (EQD) for
benzene fundamental frequencies. Degenerate frequencies are not replicated.
Values are in $cm^{-1}.$ \label{tab:benzene_results}}
\begin{tabular}{|ccc|ccc|}
\hline 
State & DC-SCIVR & EQD\textsuperscript{\citep{Halverson_Poirier_Benzene_2015}} & State & DC-SCIVR & EQD\textsuperscript{\citep{Halverson_Poirier_Benzene_2015}}\tabularnewline
\hline 
1\textsubscript{1} & 388 & 399.4554 & 10\textsubscript{1} & 1024 & 1040.98\tabularnewline
2\textsubscript{1} & 610 & 611.4227 & 11\textsubscript{1} & 1157 & 1147.751\tabularnewline
3\textsubscript{1} & 732 & 666.9294 & 12\textsubscript{1} & 1157 & 1180.374\tabularnewline
4\textsubscript{1} & 706 & 710.7318 & 13\textsubscript{1} & 1295 & 1315.612\tabularnewline
5\textsubscript{1} & 908 & 868.9106 & 14\textsubscript{1} & 1357 & 1352.563\tabularnewline
6\textsubscript{1} & 990 & 964.0127 & 15\textsubscript{1} & 1460 & 1496.231\tabularnewline
7\textsubscript{1} & 996 & 985.8294 & 16\textsubscript{1} & 1606 & 1614.455\tabularnewline
8\textsubscript{1} & 996 & 997.6235 &  &  & \tabularnewline
9\textsubscript{1} & 1018 & 1015.64 & MAE & 19 & \tabularnewline
\hline 
\end{tabular}
\par\end{centering}
\normalsize
\end{table}

\begin{figure}[!t]
\begin{centering}
\includegraphics[clip,scale=0.34]{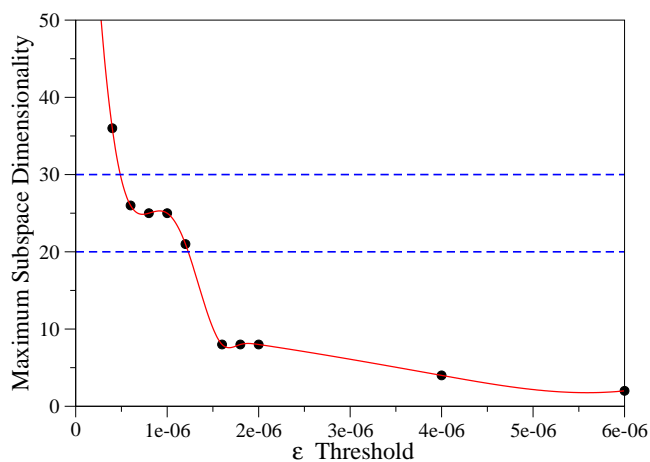}
\par\end{centering}
\caption{\label{fig:3threshold_trend}Maximum subspace dimensionality vs threshold
$\varepsilon$ for the C\protect\textsubscript{60} calculation. The
red curve fits the overall behavior, while the dashed blue lines define
the range of desired maximum subspace dimensionality. }
\end{figure}

\begin{figure}[t]
\begin{centering}
\includegraphics[scale=0.318]{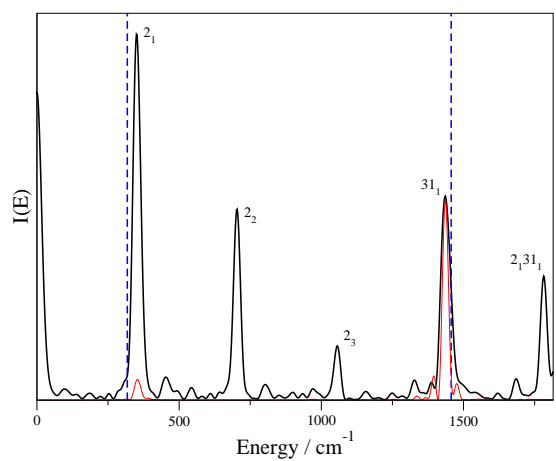}
\par\end{centering}
\caption{\label{fig:3dim_subspace}DC-SCIVR (black line) and classical (red
line) spectra for one of the subspaces employed in the C\protect\textsubscript{60}
calculation. Harmonic frequencies are reported in dashed blue lines.
Labels are according to Table (\ref{tab:fund_2nd}). }
\end{figure}

Finally, after having benchmarked the accuracy of our method against
exact quantum results for three molecules of different dimensionality
and complexity, we demonstrate applicability of DC-SCIVR to an extremely
high dimensional problem by computing the power spectrum of a fullerene-like
system. $\text{C}{}_{60}$ has 174 vibrational degrees of freedom,
a number which makes a fully quantum mechanical calculation as well
as a standard semiclassical simulation clearly unfeasible and calls
for an efficient alternative method. We employed a pre-existing force
field derived from DFT calculations on graphene sheets. This force
field takes into account stretching, bending, and torsional contributions,
but neglects bond-coupling terms and van der Waals interactions.\citep{Holec_Paris_FullereneFF_2010}
It is therefore not tailored on a real fullerene molecule, but the
main intent of this final application is to show that our method can
overcome the ``curse of dimensionality'' even in very challenging
instances. DC-SCIVR starts off with the definition of the subspaces
in which the projected spectra must be computed. Fig (\ref{fig:3threshold_trend})
shows how the choice of the threshold influences the maximum subspace
dimensionality for this system. As previously anticipated, a trade-off
leads to considering only instances within the dashed blue lines.
On the basis of Fig (\ref{fig:3threshold_trend}), we have chosen
a threshold value of $10^{-6}$, which corresponds to a maximum subspace
dimensionality equal to 25. This choice has permitted to divide the
174-dimensional vibrational space into 90 monodimensional, 1 bidimensional,
3 three-dimensional, 2 six-dimensional, 1 eight-dimensional, two 14-dimensional,
and one 25-dimensional subspaces. To calculate the projected spectra
we ran $175$ classical trajectories, each one evolved for 50000 atomic
time units. We employed a reference state $\left|\chi\right\rangle $
selected in agreement with the previously described MC-SCIVR recipe,
and, as in the case of benzene, a second-order perturbative approximation
to the pre-exponential factor. Figure (\ref{fig:3dim_subspace}) reports,
as an example, the DC-SCIVR spectrum of one of the subspaces. We have
also simulated and plotted a transient full dimensional classical
spectrum on the basis of the same trajectories employed for the semiclassical
calculations. To better compare the two different simulations we have
shifted the DC-SCIVR spectrum in such a way that the zero-point energy
is set to zero. From the comparison, we note that DC-SCIVR and classical
estimates are close to each other. However, DC-SCIVR is able to increase
the level of knowledge by detecting also quantum overtones. Results
up to an energy of about 1600 cm\textsuperscript{-1} relative to
the zero point energy can be found in Table (\ref{tab:fund_2nd}). 

\onecolumngrid

\begin{table}[!b]
\begin{centering}
\footnotesize\caption{\label{tab:fund_2nd}Frequencies of the C\protect\textsubscript{60}
model up to 1600$cm^{-1}$. ``HO'' indicates harmonic values; ``Cl''
labels the classical estimates of fundamental frequencies; ``DC-SCIVR''
introduces our semiclassical results. Values are in $cm^{-1}.$ }
\par\end{centering}
\begin{centering}
\begin{tabular}{|cccc|cccc|cccc|cccc|cccc|}
\hline 
State & HO & Cl & DC-SCIVR & St. & HO & Cl & DC-SCIVR & St. & HO & Cl & DC-SCIVR & St. & HO & Cl & DC-SCIVR & St. & HO & Cl & DC-SCIVR\tabularnewline
\hline 
$1_{1}$ & 255 & 254 & 254 & $10_{1}$ & 568 & 611 & 610 & $5_{2}$ & 808 &  & 807 & $23_{1}$ & 1014 & 1015 & 1015 & $29_{1}$ & 1310 & 1269 & 1264\tabularnewline
$2_{1}$ & 318 & 355 & 352 & $11_{1}$ & 601 & 571 & 572 & $17_{1}$ & 816 & 779 & 774 & $24_{1}$ & 1042 & 1039 & 1037 & $13_{2}$ & 1314 &  & 1303\tabularnewline
$3_{1}$ & 359 & 347 & 346 & $2_{2}$ & 636 &  & 706 & $18_{1}$ & 863 & 872 & 872 & $25_{1}$ & 1052 & 1075 & 1075 & $31_{1}$ & 1457 & 1438 & 1434\tabularnewline
$4_{1}$ & 404 & 432 & 432 & $12_{1}$ & 648 & 630 & 626 & $19_{1}$ & 890 & 911 & 911 & $26_{1}$ & 1091 & 1062 & 1060 & $32_{1}$ & 1470 & 1398 & 1391\tabularnewline
$5_{1}$ & 404 & 404 & 403 & $13_{1}$ & 657 & 652 & 651 & $20_{1}$ & 905 & 880 & 880 & $9_{2}$ & 1092 &  & 1091 & $33_{1}$ & 1526 & 1467 & 1506\tabularnewline
$6_{1}$ & 484 & 483 & 483 & $3_{2}$ & 718 &  & 693 & $21_{1}$ & 962 & 971 & 971 & $10_{2}$ & 1136 &  & 1220 & $14_{2}$ & 1540 &  & 1534\tabularnewline
$7_{1}$ & 488 & 547 & 547 & $14_{1}$ & 770 & 767 & 767 & $6_{2}$ & 968 &  & 966 & $11_{2}$ & 1202 &  & 1150 & $15_{2}$ & 1550 &  & 1533\tabularnewline
$8_{1}$ & 494 & 478 & 478 & $15_{1}$ & 775 & 766 & 766 & $7_{2}$ & 976 &  & 1093 & $27_{1}$ & 1225 & 1218 & 1218 & $16_{2}$ & 1562 &  & 1554\tabularnewline
$1_{2}$ & 510 &  & 506 & $16_{1}$ & 781 & 777 & 777 & $8_{2}$ & 988 &  & 957 & $28_{1}$ & 1252  & 1231 & 1231 &  &  &  & \tabularnewline
$9_{1}$ & 546 & 545 & 546 & $4_{2}$ & 808 &  & 863 & $22_{1}$ & 1000 & 997 & 998 & $12_{2}$ & 1296 &  & 1254 &  &  &  & \tabularnewline
\hline 
\end{tabular}
\par\end{centering}
\normalsize
\end{table}

\newpage  \twocolumngrid A concern that may arise about the approach
regards its efficiency when dealing with lower-symmetry molecules.
Thus, to demonstrate that reduced symmetry is not a hindrance to our
calculations, we have investigated an \textit{ad hoc} constructed
fullerene isotope model for which symmetry has been broken. Substitution
of three appropriate carbon nuclei with nuclei having the same mass
of gold ones removed the degeneracies of the vibrational levels. This
model was built to preserve the original nuclear and electronic charges,
so that the force field did not need to be modified.\\The result
of the isotopic substitution is that previously degenerate frequencies
are split already at the harmonic level. Even if such splittings are
mostly within semiclassical accuracy (i.e. 25-30 cm\textsuperscript{-1}),
DC-SCIVR results are resolved enough to detect a multiple-peak feature
in the isotopic model opposite to the original case characterized
by a lonely (degenerate) peak. A relevant example of this is reported
in the Supplemental Material.\citep{Suppl_Info_PRL}

In summary, we have presented a new approach to the calculation of
theoretical vibrational spectra of high dimensional molecular systems.
The method has been tested for the small and highly inter-mode coupled
water molecule, the highly chaotic methane molecule, and the high
dimensional benzene molecule yielding in all cases accurate estimates
if compared to available exact quantum results. Then, application
to a sizable system made of 174 degrees of freedom has demonstrated
that even for such large systems an accurate quantum estimate of fundamental
and overtone frequencies is feasible, thus opening up the possibility
to quantum investigate the spectroscopy of highly dimensional systems. 
\begin{acknowledgments}
We acknowledge support from the European Research Council (ERC) under
the European Union\textquoteright s Horizon 2020 research and innovation
programme (grant agreement No {[}647107{]} \textendash{} SEMICOMPLEX
\textendash{} ERC-2014-CoG).
\end{acknowledgments}

\bibliographystyle{apsrev4-1}
\bibliography{SEMICOMPLEX}

\begin{thebibliography}{91}%
\makeatletter
\providecommand \@ifxundefined [1]{%
 \@ifx{#1\undefined}
}%
\providecommand \@ifnum [1]{%
 \ifnum #1\expandafter \@firstoftwo
 \else \expandafter \@secondoftwo
 \fi
}%
\providecommand \@ifx [1]{%
 \ifx #1\expandafter \@firstoftwo
 \else \expandafter \@secondoftwo
 \fi
}%
\providecommand \natexlab [1]{#1}%
\providecommand \enquote  [1]{``#1''}%
\providecommand \bibnamefont  [1]{#1}%
\providecommand \bibfnamefont [1]{#1}%
\providecommand \citenamefont [1]{#1}%
\providecommand \href@noop [0]{\@secondoftwo}%
\providecommand \href [0]{\begingroup \@sanitize@url \@href}%
\providecommand \@href[1]{\@@startlink{#1}\@@href}%
\providecommand \@@href[1]{\endgroup#1\@@endlink}%
\providecommand \@sanitize@url [0]{\catcode `\\12\catcode `\$12\catcode
  `\&12\catcode `\#12\catcode `\^12\catcode `\_12\catcode `\%12\relax}%
\providecommand \@@startlink[1]{}%
\providecommand \@@endlink[0]{}%
\providecommand \url  [0]{\begingroup\@sanitize@url \@url }%
\providecommand \@url [1]{\endgroup\@href {#1}{\urlprefix }}%
\providecommand \urlprefix  [0]{URL }%
\providecommand \Eprint [0]{\href }%
\providecommand \doibase [0]{http://dx.doi.org/}%
\providecommand \selectlanguage [0]{\@gobble}%
\providecommand \bibinfo  [0]{\@secondoftwo}%
\providecommand \bibfield  [0]{\@secondoftwo}%
\providecommand \translation [1]{[#1]}%
\providecommand \BibitemOpen [0]{}%
\providecommand \bibitemStop [0]{}%
\providecommand \bibitemNoStop [0]{.\EOS\space}%
\providecommand \EOS [0]{\spacefactor3000\relax}%
\providecommand \BibitemShut  [1]{\csname bibitem#1\endcsname}%
\let\auto@bib@innerbib\@empty
%</preamble>
\bibitem [{\citenamefont {Bowman}\ \emph {et~al.}(2003)\citenamefont {Bowman},
  \citenamefont {Carter},\ and\ \citenamefont
  {Huang}}]{Bowman_Huang_MULTIMODE_2003}%
  \BibitemOpen
  \bibfield  {author} {\bibinfo {author} {\bibfnamefont {J.~M.}\ \bibnamefont
  {Bowman}}, \bibinfo {author} {\bibfnamefont {S.}~\bibnamefont {Carter}}, \
  and\ \bibinfo {author} {\bibfnamefont {X.}~\bibnamefont {Huang}},\
  }\href@noop {} {\bibfield  {journal} {\bibinfo  {journal} {Int. Rev. Phys.
  Chem.}\ }\textbf {\bibinfo {volume} {22}},\ \bibinfo {pages} {533} (\bibinfo
  {year} {2003})}\BibitemShut {NoStop}%
\bibitem [{\citenamefont {Samanta}\ \emph {et~al.}(2016)\citenamefont
  {Samanta}, \citenamefont {Wang}, \citenamefont {Mancini}, \citenamefont
  {Bowman},\ and\ \citenamefont
  {Reisler}}]{Samanta_Reisler_Mixedclusters_2016}%
  \BibitemOpen
  \bibfield  {author} {\bibinfo {author} {\bibfnamefont {A.~K.}\ \bibnamefont
  {Samanta}}, \bibinfo {author} {\bibfnamefont {Y.}~\bibnamefont {Wang}},
  \bibinfo {author} {\bibfnamefont {J.~S.}\ \bibnamefont {Mancini}}, \bibinfo
  {author} {\bibfnamefont {J.~M.}\ \bibnamefont {Bowman}}, \ and\ \bibinfo
  {author} {\bibfnamefont {H.}~\bibnamefont {Reisler}},\ }\href {\doibase
  10.1021/acs.chemrev.5b00506} {\bibfield  {journal} {\bibinfo  {journal}
  {Chem. Rev.}\ }\textbf {\bibinfo {volume} {116}},\ \bibinfo {pages} {4913}
  (\bibinfo {year} {2016})}\BibitemShut {NoStop}%
\bibitem [{\citenamefont {Qu}\ \emph {et~al.}(2015)\citenamefont {Qu},
  \citenamefont {Conte}, \citenamefont {Houston},\ and\ \citenamefont
  {Bowman}}]{Chen_Bowman_Methane-Water_2015}%
  \BibitemOpen
  \bibfield  {author} {\bibinfo {author} {\bibfnamefont {C.}~\bibnamefont
  {Qu}}, \bibinfo {author} {\bibfnamefont {R.}~\bibnamefont {Conte}}, \bibinfo
  {author} {\bibfnamefont {P.~L.}\ \bibnamefont {Houston}}, \ and\ \bibinfo
  {author} {\bibfnamefont {J.~M.}\ \bibnamefont {Bowman}},\ }\href@noop {}
  {\bibfield  {journal} {\bibinfo  {journal} {Phys. Chem. Chem. Phys.}\
  }\textbf {\bibinfo {volume} {17}},\ \bibinfo {pages} {8172} (\bibinfo {year}
  {2015})}\BibitemShut {NoStop}%
\bibitem [{\citenamefont {Avila}\ and\ \citenamefont
  {Carrington~Jr}(2011)}]{Avila_Carrington_C2H4_2011}%
  \BibitemOpen
  \bibfield  {author} {\bibinfo {author} {\bibfnamefont {G.}~\bibnamefont
  {Avila}}\ and\ \bibinfo {author} {\bibfnamefont {T.}~\bibnamefont
  {Carrington~Jr}},\ }\href@noop {} {\bibfield  {journal} {\bibinfo  {journal}
  {J. Chem. Phys.}\ }\textbf {\bibinfo {volume} {135}},\ \bibinfo {pages}
  {064101} (\bibinfo {year} {2011})}\BibitemShut {NoStop}%
\bibitem [{\citenamefont {Manthe}(2002)}]{manthe2002reaction}%
  \BibitemOpen
  \bibfield  {author} {\bibinfo {author} {\bibfnamefont {U.}~\bibnamefont
  {Manthe}},\ }\href@noop {} {\bibfield  {journal} {\bibinfo  {journal} {J.
  Theor. Comp. Chem.}\ }\textbf {\bibinfo {volume} {1}},\ \bibinfo {pages}
  {153} (\bibinfo {year} {2002})}\BibitemShut {NoStop}%
\bibitem [{\citenamefont {Meyer}\ and\ \citenamefont
  {Worth}(2003)}]{Meyer_Worth_HighDimMCTDH_2003}%
  \BibitemOpen
  \bibfield  {author} {\bibinfo {author} {\bibfnamefont {H.-D.}\ \bibnamefont
  {Meyer}}\ and\ \bibinfo {author} {\bibfnamefont {G.~A.}\ \bibnamefont
  {Worth}},\ }\href@noop {} {\bibfield  {journal} {\bibinfo  {journal} {Theor.
  Chem. Acc.}\ }\textbf {\bibinfo {volume} {109}},\ \bibinfo {pages} {251}
  (\bibinfo {year} {2003})}\BibitemShut {NoStop}%
\bibitem [{\citenamefont {Bowman}\ \emph {et~al.}(2008)\citenamefont {Bowman},
  \citenamefont {Carrington},\ and\ \citenamefont
  {Meyer}}]{Bowman_Meyer_Polyatomic_2008}%
  \BibitemOpen
  \bibfield  {author} {\bibinfo {author} {\bibfnamefont {J.~M.}\ \bibnamefont
  {Bowman}}, \bibinfo {author} {\bibfnamefont {T.}~\bibnamefont {Carrington}},
  \ and\ \bibinfo {author} {\bibfnamefont {H.-D.}\ \bibnamefont {Meyer}},\
  }\href@noop {} {\bibfield  {journal} {\bibinfo  {journal} {Molecular
  Physics}\ }\textbf {\bibinfo {volume} {106}},\ \bibinfo {pages} {2145}
  (\bibinfo {year} {2008})}\BibitemShut {NoStop}%
\bibitem [{\citenamefont {Bludsky}\ \emph {et~al.}(2000)\citenamefont
  {Bludsky}, \citenamefont {Chocholousova}, \citenamefont {Vacek},
  \citenamefont {Huisken},\ and\ \citenamefont
  {Hobza}}]{Bludsky_Hobza_Anharmonicgly_2000}%
  \BibitemOpen
  \bibfield  {author} {\bibinfo {author} {\bibfnamefont {O.}~\bibnamefont
  {Bludsky}}, \bibinfo {author} {\bibfnamefont {J.}~\bibnamefont
  {Chocholousova}}, \bibinfo {author} {\bibfnamefont {J.}~\bibnamefont
  {Vacek}}, \bibinfo {author} {\bibfnamefont {F.}~\bibnamefont {Huisken}}, \
  and\ \bibinfo {author} {\bibfnamefont {P.}~\bibnamefont {Hobza}},\ }\href
  {\doibase http://dx.doi.org/10.1063/1.1288914} {\bibfield  {journal}
  {\bibinfo  {journal} {J. Chem. Phys.}\ }\textbf {\bibinfo {volume} {113}},\
  \bibinfo {pages} {4629} (\bibinfo {year} {2000})}\BibitemShut {NoStop}%
\bibitem [{\citenamefont {Barone}(2005)}]{Barone_AutomatedVPT2_2005}%
  \BibitemOpen
  \bibfield  {author} {\bibinfo {author} {\bibfnamefont {V.}~\bibnamefont
  {Barone}},\ }\href {\doibase http://dx.doi.org/10.1063/1.1824881} {\bibfield
  {journal} {\bibinfo  {journal} {J. Chem. Phys.}\ }\textbf {\bibinfo {volume}
  {122}},\ \bibinfo {pages} {014108} (\bibinfo {year} {2005})}\BibitemShut
  {NoStop}%
\bibitem [{\citenamefont {Biczysko}\ \emph {et~al.}(2012)\citenamefont
  {Biczysko}, \citenamefont {Bloino}, \citenamefont {Carnimeo}, \citenamefont
  {Panek},\ and\ \citenamefont {Barone}}]{Biczysko_Barone_abinitioIRgly_2012}%
  \BibitemOpen
  \bibfield  {author} {\bibinfo {author} {\bibfnamefont {M.}~\bibnamefont
  {Biczysko}}, \bibinfo {author} {\bibfnamefont {J.}~\bibnamefont {Bloino}},
  \bibinfo {author} {\bibfnamefont {I.}~\bibnamefont {Carnimeo}}, \bibinfo
  {author} {\bibfnamefont {P.}~\bibnamefont {Panek}}, \ and\ \bibinfo {author}
  {\bibfnamefont {V.}~\bibnamefont {Barone}},\ }\href {\doibase
  http://dx.doi.org/10.1016/j.molstruc.2011.10.012} {\bibfield  {journal}
  {\bibinfo  {journal} {J. Mol. Struct.}\ }\textbf {\bibinfo {volume} {1009}},\
  \bibinfo {pages} {74 } (\bibinfo {year} {2012})}\BibitemShut {NoStop}%
\bibitem [{\citenamefont {Elran}\ and\ \citenamefont
  {Kay}(1999)}]{Elran_Kay_ImprovingHK_1999}%
  \BibitemOpen
  \bibfield  {author} {\bibinfo {author} {\bibfnamefont {Y.}~\bibnamefont
  {Elran}}\ and\ \bibinfo {author} {\bibfnamefont {K.}~\bibnamefont {Kay}},\
  }\href@noop {} {\bibfield  {journal} {\bibinfo  {journal} {J. Chem. Phys.}\
  }\textbf {\bibinfo {volume} {110}},\ \bibinfo {pages} {3653} (\bibinfo {year}
  {1999})}\BibitemShut {NoStop}%
\bibitem [{\citenamefont {Zhang}\ and\ \citenamefont
  {Pollak}(2004)}]{Zhang_Pollak_Deeptunneling_2004}%
  \BibitemOpen
  \bibfield  {author} {\bibinfo {author} {\bibfnamefont {D.~H.}\ \bibnamefont
  {Zhang}}\ and\ \bibinfo {author} {\bibfnamefont {E.}~\bibnamefont {Pollak}},\
  }\href@noop {} {\bibfield  {journal} {\bibinfo  {journal} {Phys. Rev. Lett.}\
  }\textbf {\bibinfo {volume} {93}},\ \bibinfo {pages} {140401} (\bibinfo
  {year} {2004})}\BibitemShut {NoStop}%
\bibitem [{\citenamefont {Kay}(2006)}]{Kay_SCcorrections_2006}%
  \BibitemOpen
  \bibfield  {author} {\bibinfo {author} {\bibfnamefont {K.~G.}\ \bibnamefont
  {Kay}},\ }\href {\doibase http://dx.doi.org/10.1016/j.chemphys.2005.06.019}
  {\bibfield  {journal} {\bibinfo  {journal} {Chem. Phys.}\ }\textbf {\bibinfo
  {volume} {322}},\ \bibinfo {pages} {3 } (\bibinfo {year} {2006})}\BibitemShut
  {NoStop}%
\bibitem [{\citenamefont {Conte}\ and\ \citenamefont
  {Pollak}(2010)}]{Conte_Pollak_ThawedGaussian_2010}%
  \BibitemOpen
  \bibfield  {author} {\bibinfo {author} {\bibfnamefont {R.}~\bibnamefont
  {Conte}}\ and\ \bibinfo {author} {\bibfnamefont {E.}~\bibnamefont {Pollak}},\
  }\href@noop {} {\bibfield  {journal} {\bibinfo  {journal} {Phys. Rev. E}\
  }\textbf {\bibinfo {volume} {81}},\ \bibinfo {pages} {036704} (\bibinfo
  {year} {2010})}\BibitemShut {NoStop}%
\bibitem [{\citenamefont {Bonella}\ \emph {et~al.}(2010)\citenamefont
  {Bonella}, \citenamefont {Ciccotti},\ and\ \citenamefont
  {Kapral}}]{Bonella_Kapral_quantum-classical_2010}%
  \BibitemOpen
  \bibfield  {author} {\bibinfo {author} {\bibfnamefont {S.}~\bibnamefont
  {Bonella}}, \bibinfo {author} {\bibfnamefont {G.}~\bibnamefont {Ciccotti}}, \
  and\ \bibinfo {author} {\bibfnamefont {R.}~\bibnamefont {Kapral}},\
  }\href@noop {} {\bibfield  {journal} {\bibinfo  {journal} {Chem. Phys.
  Lett.}\ }\textbf {\bibinfo {volume} {484}},\ \bibinfo {pages} {399} (\bibinfo
  {year} {2010})}\BibitemShut {NoStop}%
\bibitem [{\citenamefont {Monteferrante}\ \emph {et~al.}(2013)\citenamefont
  {Monteferrante}, \citenamefont {Bonella},\ and\ \citenamefont
  {Ciccotti}}]{Monteferrante_Ciccotti_Liquidneon_2013}%
  \BibitemOpen
  \bibfield  {author} {\bibinfo {author} {\bibfnamefont {M.}~\bibnamefont
  {Monteferrante}}, \bibinfo {author} {\bibfnamefont {S.}~\bibnamefont
  {Bonella}}, \ and\ \bibinfo {author} {\bibfnamefont {G.}~\bibnamefont
  {Ciccotti}},\ }\href@noop {} {\bibfield  {journal} {\bibinfo  {journal} {J.
  Chem. Phys.}\ }\textbf {\bibinfo {volume} {138}},\ \bibinfo {pages} {054118}
  (\bibinfo {year} {2013})}\BibitemShut {NoStop}%
\bibitem [{\citenamefont {Conte}\ and\ \citenamefont
  {Pollak}(2012)}]{Conte_Pollak_ContinuumLimit_2012}%
  \BibitemOpen
  \bibfield  {author} {\bibinfo {author} {\bibfnamefont {R.}~\bibnamefont
  {Conte}}\ and\ \bibinfo {author} {\bibfnamefont {E.}~\bibnamefont {Pollak}},\
  }\href@noop {} {\bibfield  {journal} {\bibinfo  {journal} {J. Chem. Phys.}\
  }\textbf {\bibinfo {volume} {136}},\ \bibinfo {pages} {094101} (\bibinfo
  {year} {2012})}\BibitemShut {NoStop}%
\bibitem [{\citenamefont {Petersen}\ and\ \citenamefont
  {Pollak}(2015)}]{Petersen_Pollak_Interactionrepresentation_2015}%
  \BibitemOpen
  \bibfield  {author} {\bibinfo {author} {\bibfnamefont {J.}~\bibnamefont
  {Petersen}}\ and\ \bibinfo {author} {\bibfnamefont {E.}~\bibnamefont
  {Pollak}},\ }\href@noop {} {\bibfield  {journal} {\bibinfo  {journal} {J.
  Chem. Phys.}\ }\textbf {\bibinfo {volume} {143}},\ \bibinfo {pages} {224114}
  (\bibinfo {year} {2015})}\BibitemShut {NoStop}%
\bibitem [{\citenamefont {Shalashilin}\ and\ \citenamefont
  {Child}(2001)}]{Shalashilin_Child_Coherentstates_2001}%
  \BibitemOpen
  \bibfield  {author} {\bibinfo {author} {\bibfnamefont {D.~V.}\ \bibnamefont
  {Shalashilin}}\ and\ \bibinfo {author} {\bibfnamefont {M.~S.}\ \bibnamefont
  {Child}},\ }\href@noop {} {\bibfield  {journal} {\bibinfo  {journal} {J.
  Chem. Phys.}\ }\textbf {\bibinfo {volume} {115}},\ \bibinfo {pages} {5367}
  (\bibinfo {year} {2001})}\BibitemShut {NoStop}%
\bibitem [{\citenamefont {Shalashilin}\ and\ \citenamefont
  {Child}(2004)}]{Shalashilin_Child_CCS_2004}%
  \BibitemOpen
  \bibfield  {author} {\bibinfo {author} {\bibfnamefont {D.~V.}\ \bibnamefont
  {Shalashilin}}\ and\ \bibinfo {author} {\bibfnamefont {M.~S.}\ \bibnamefont
  {Child}},\ }\href@noop {} {\bibfield  {journal} {\bibinfo  {journal} {Chem.
  Phys.}\ }\textbf {\bibinfo {volume} {304}},\ \bibinfo {pages} {103} (\bibinfo
  {year} {2004})}\BibitemShut {NoStop}%
\bibitem [{\citenamefont {Heatwole}\ and\ \citenamefont
  {Prezhdo}(2009)}]{Heatwole_Prezhdo_Morsesemiclassical_2009}%
  \BibitemOpen
  \bibfield  {author} {\bibinfo {author} {\bibfnamefont {E.~M.}\ \bibnamefont
  {Heatwole}}\ and\ \bibinfo {author} {\bibfnamefont {O.~V.}\ \bibnamefont
  {Prezhdo}},\ }\href@noop {} {\bibfield  {journal} {\bibinfo  {journal} {J.
  Chem. Phys.}\ }\textbf {\bibinfo {volume} {130}},\ \bibinfo {pages} {244111}
  (\bibinfo {year} {2009})}\BibitemShut {NoStop}%
\bibitem [{\citenamefont {Garashchuk}\ \emph {et~al.}(2011)\citenamefont
  {Garashchuk}, \citenamefont {Rassolov},\ and\ \citenamefont
  {Prezhdo}}]{Garashchuk_Prezhdo_Bohmian_2011}%
  \BibitemOpen
  \bibfield  {author} {\bibinfo {author} {\bibfnamefont {S.}~\bibnamefont
  {Garashchuk}}, \bibinfo {author} {\bibfnamefont {V.}~\bibnamefont
  {Rassolov}}, \ and\ \bibinfo {author} {\bibfnamefont {O.}~\bibnamefont
  {Prezhdo}},\ }\href@noop {} {\bibfield  {journal} {\bibinfo  {journal} {Rev.
  Comput. Chem.}\ }\textbf {\bibinfo {volume} {27}},\ \bibinfo {pages} {287}
  (\bibinfo {year} {2011})}\BibitemShut {NoStop}%
\bibitem [{\citenamefont {Huo}\ and\ \citenamefont
  {Coker}(2012)}]{Huo_Coker_Semiclassicalnonadiabatic_2012}%
  \BibitemOpen
  \bibfield  {author} {\bibinfo {author} {\bibfnamefont {P.}~\bibnamefont
  {Huo}}\ and\ \bibinfo {author} {\bibfnamefont {D.~F.}\ \bibnamefont
  {Coker}},\ }\href@noop {} {\bibfield  {journal} {\bibinfo  {journal} {Mol.
  Phys.}\ }\textbf {\bibinfo {volume} {110}},\ \bibinfo {pages} {1035}
  (\bibinfo {year} {2012})}\BibitemShut {NoStop}%
\bibitem [{\citenamefont {Pal}\ \emph {et~al.}(2016)\citenamefont {Pal},
  \citenamefont {Vyas},\ and\ \citenamefont
  {Tomsovic}}]{Harinder_Tomsovic_GeneralizedGaussian_2016}%
  \BibitemOpen
  \bibfield  {author} {\bibinfo {author} {\bibfnamefont {H.}~\bibnamefont
  {Pal}}, \bibinfo {author} {\bibfnamefont {M.}~\bibnamefont {Vyas}}, \ and\
  \bibinfo {author} {\bibfnamefont {S.}~\bibnamefont {Tomsovic}},\ }\href
  {\doibase 10.1103/PhysRevE.93.012213} {\bibfield  {journal} {\bibinfo
  {journal} {Phys. Rev. E}\ }\textbf {\bibinfo {volume} {93}},\ \bibinfo
  {pages} {012213} (\bibinfo {year} {2016})}\BibitemShut {NoStop}%
\bibitem [{\citenamefont {Koch}\ and\ \citenamefont
  {Tannor}(2017)}]{Koch_Tannor_WPrevival_2017}%
  \BibitemOpen
  \bibfield  {author} {\bibinfo {author} {\bibfnamefont {W.}~\bibnamefont
  {Koch}}\ and\ \bibinfo {author} {\bibfnamefont {D.~J.}\ \bibnamefont
  {Tannor}},\ }\href@noop {} {\bibfield  {journal} {\bibinfo  {journal}
  {arXiv:1701.01378}\ } (\bibinfo {year} {2017})}\BibitemShut {NoStop}%
\bibitem [{\citenamefont {Harabati}\ \emph {et~al.}(2004)\citenamefont
  {Harabati}, \citenamefont {Rost},\ and\ \citenamefont
  {Grossmann}}]{Harabati_Grossmann_LongtimeSCIVR_2004}%
  \BibitemOpen
  \bibfield  {author} {\bibinfo {author} {\bibfnamefont {C.}~\bibnamefont
  {Harabati}}, \bibinfo {author} {\bibfnamefont {J.~M.}\ \bibnamefont {Rost}},
  \ and\ \bibinfo {author} {\bibfnamefont {F.}~\bibnamefont {Grossmann}},\
  }\href@noop {} {\bibfield  {journal} {\bibinfo  {journal} {J. Chem. Phys.}\
  }\textbf {\bibinfo {volume} {120}},\ \bibinfo {pages} {26} (\bibinfo {year}
  {2004})}\BibitemShut {NoStop}%
\bibitem [{\citenamefont {Grossmann}(1999)}]{Grossmann_HierarchySC_1999}%
  \BibitemOpen
  \bibfield  {author} {\bibinfo {author} {\bibfnamefont {F.}~\bibnamefont
  {Grossmann}},\ }\href@noop {} {\bibfield  {journal} {\bibinfo  {journal}
  {Comments At. Mol. Phys.}\ }\textbf {\bibinfo {volume} {34}},\ \bibinfo
  {pages} {141} (\bibinfo {year} {1999})}\BibitemShut {NoStop}%
\bibitem [{\citenamefont {Nakamura}\ \emph {et~al.}(2016)\citenamefont
  {Nakamura}, \citenamefont {Nanbu}, \citenamefont {Teranishi},\ and\
  \citenamefont {Ohta}}]{Nakamura_Ohta_SCDevelopment_2016}%
  \BibitemOpen
  \bibfield  {author} {\bibinfo {author} {\bibfnamefont {H.}~\bibnamefont
  {Nakamura}}, \bibinfo {author} {\bibfnamefont {S.}~\bibnamefont {Nanbu}},
  \bibinfo {author} {\bibfnamefont {Y.}~\bibnamefont {Teranishi}}, \ and\
  \bibinfo {author} {\bibfnamefont {A.}~\bibnamefont {Ohta}},\ }\href@noop {}
  {\bibfield  {journal} {\bibinfo  {journal} {Phys. Chem. Chem. Phys.}\
  }\textbf {\bibinfo {volume} {18}},\ \bibinfo {pages} {11972} (\bibinfo {year}
  {2016})}\BibitemShut {NoStop}%
\bibitem [{\citenamefont {Kondorskiy}\ and\ \citenamefont
  {Nanbu}(2015)}]{Kondorskiy_Nanbu_Nonadiabatic_2015}%
  \BibitemOpen
  \bibfield  {author} {\bibinfo {author} {\bibfnamefont {A.~D.}\ \bibnamefont
  {Kondorskiy}}\ and\ \bibinfo {author} {\bibfnamefont {S.}~\bibnamefont
  {Nanbu}},\ }\href@noop {} {\bibfield  {journal} {\bibinfo  {journal} {J.
  Chem. Phys.}\ }\textbf {\bibinfo {volume} {143}},\ \bibinfo {pages} {114103}
  (\bibinfo {year} {2015})}\BibitemShut {NoStop}%
\bibitem [{\citenamefont {Tao}(2014)}]{Tao_ImportanceSampling_2014}%
  \BibitemOpen
  \bibfield  {author} {\bibinfo {author} {\bibfnamefont {G.}~\bibnamefont
  {Tao}},\ }\href@noop {} {\bibfield  {journal} {\bibinfo  {journal} {Theor.
  Chem. Acc.}\ }\textbf {\bibinfo {volume} {133}},\ \bibinfo {pages} {1448}
  (\bibinfo {year} {2014})}\BibitemShut {NoStop}%
\bibitem [{\citenamefont {Antipov}\ \emph {et~al.}(2015)\citenamefont
  {Antipov}, \citenamefont {Ye},\ and\ \citenamefont
  {Ananth}}]{Antipov_Nandini_Mixedqcl_2015}%
  \BibitemOpen
  \bibfield  {author} {\bibinfo {author} {\bibfnamefont {S.~V.}\ \bibnamefont
  {Antipov}}, \bibinfo {author} {\bibfnamefont {Z.}~\bibnamefont {Ye}}, \ and\
  \bibinfo {author} {\bibfnamefont {N.}~\bibnamefont {Ananth}},\ }\href
  {\doibase http://dx.doi.org/10.1063/1.4919667} {\bibfield  {journal}
  {\bibinfo  {journal} {J. Chem. Phys.}\ }\textbf {\bibinfo {volume} {142}},\
  \bibinfo {eid} {184102} (\bibinfo {year} {2015}),\
  http://dx.doi.org/10.1063/1.4919667}\BibitemShut {NoStop}%
\bibitem [{\citenamefont {Liu}\ and\ \citenamefont
  {Miller}(2006)}]{Liu_Miller_ThermalGaussian_2006}%
  \BibitemOpen
  \bibfield  {author} {\bibinfo {author} {\bibfnamefont {J.}~\bibnamefont
  {Liu}}\ and\ \bibinfo {author} {\bibfnamefont {W.~H.}\ \bibnamefont
  {Miller}},\ }\href@noop {} {\bibfield  {journal} {\bibinfo  {journal} {J.
  Chem. Phys.}\ }\textbf {\bibinfo {volume} {125}},\ \bibinfo {pages} {224104}
  (\bibinfo {year} {2006})}\BibitemShut {NoStop}%
\bibitem [{\citenamefont {Liu}\ and\ \citenamefont
  {Miller}(2007{\natexlab{a}})}]{Liu_Miller_linearizedSCIVR_2007}%
  \BibitemOpen
  \bibfield  {author} {\bibinfo {author} {\bibfnamefont {J.}~\bibnamefont
  {Liu}}\ and\ \bibinfo {author} {\bibfnamefont {W.~H.}\ \bibnamefont
  {Miller}},\ }\href@noop {} {\bibfield  {journal} {\bibinfo  {journal} {J.
  Chem. Phys.}\ }\textbf {\bibinfo {volume} {127}},\ \bibinfo {pages} {114506}
  (\bibinfo {year} {2007}{\natexlab{a}})}\BibitemShut {NoStop}%
\bibitem [{\citenamefont {Liu}\ and\ \citenamefont
  {Miller}(2007{\natexlab{b}})}]{Liu_Miller_RealTimecorrelation_2007}%
  \BibitemOpen
  \bibfield  {author} {\bibinfo {author} {\bibfnamefont {J.}~\bibnamefont
  {Liu}}\ and\ \bibinfo {author} {\bibfnamefont {W.~H.}\ \bibnamefont
  {Miller}},\ }\href@noop {} {\bibfield  {journal} {\bibinfo  {journal} {J.
  Chem. Phys.}\ }\textbf {\bibinfo {volume} {126}},\ \bibinfo {pages} {234110}
  (\bibinfo {year} {2007}{\natexlab{b}})}\BibitemShut {NoStop}%
\bibitem [{\citenamefont {Liu}\ and\ \citenamefont
  {Miller}(2008)}]{Liu_Miller_VariouslinearizedSCIVR_2008}%
  \BibitemOpen
  \bibfield  {author} {\bibinfo {author} {\bibfnamefont {J.}~\bibnamefont
  {Liu}}\ and\ \bibinfo {author} {\bibfnamefont {W.~H.}\ \bibnamefont
  {Miller}},\ }\href@noop {} {\bibfield  {journal} {\bibinfo  {journal} {J.
  Chem. Phys.}\ }\textbf {\bibinfo {volume} {128}},\ \bibinfo {pages} {144511}
  (\bibinfo {year} {2008})}\BibitemShut {NoStop}%
\bibitem [{\citenamefont {Koda}(2015)}]{Koda_SCIVRWigner_2015}%
  \BibitemOpen
  \bibfield  {author} {\bibinfo {author} {\bibfnamefont {S.-I.}\ \bibnamefont
  {Koda}},\ }\href@noop {} {\bibfield  {journal} {\bibinfo  {journal} {J. Chem.
  Phys.}\ }\textbf {\bibinfo {volume} {143}},\ \bibinfo {pages} {244110}
  (\bibinfo {year} {2015})}\BibitemShut {NoStop}%
\bibitem [{\citenamefont {Koda}(2016)}]{Koda_Mixedsemiclassical_2016}%
  \BibitemOpen
  \bibfield  {author} {\bibinfo {author} {\bibfnamefont {S.-I.}\ \bibnamefont
  {Koda}},\ }\href@noop {} {\bibfield  {journal} {\bibinfo  {journal} {J. Chem.
  Phys.}\ }\textbf {\bibinfo {volume} {144}},\ \bibinfo {pages} {154108}
  (\bibinfo {year} {2016})}\BibitemShut {NoStop}%
\bibitem [{\citenamefont {Ushiyama}\ and\ \citenamefont
  {Takatsuka}(2005)}]{Ushiyama_Takatsuka_Interferences_2005}%
  \BibitemOpen
  \bibfield  {author} {\bibinfo {author} {\bibfnamefont {H.}~\bibnamefont
  {Ushiyama}}\ and\ \bibinfo {author} {\bibfnamefont {K.}~\bibnamefont
  {Takatsuka}},\ }\href@noop {} {\bibfield  {journal} {\bibinfo  {journal} {J.
  Chem. Phys.}\ }\textbf {\bibinfo {volume} {122}},\ \bibinfo {pages} {224112}
  (\bibinfo {year} {2005})}\BibitemShut {NoStop}%
\bibitem [{\citenamefont {Takahashi}\ and\ \citenamefont
  {Takatsuka}(2007)}]{Takahashi_Takatsuka_PhaseQuantization_2007}%
  \BibitemOpen
  \bibfield  {author} {\bibinfo {author} {\bibfnamefont {S.}~\bibnamefont
  {Takahashi}}\ and\ \bibinfo {author} {\bibfnamefont {K.}~\bibnamefont
  {Takatsuka}},\ }\href@noop {} {\bibfield  {journal} {\bibinfo  {journal} {J.
  Chem. Phys.}\ }\textbf {\bibinfo {volume} {127}},\ \bibinfo {pages} {084112}
  (\bibinfo {year} {2007})}\BibitemShut {NoStop}%
\bibitem [{\citenamefont {Zhuang}\ \emph {et~al.}(2012)\citenamefont {Zhuang},
  \citenamefont {Siebert}, \citenamefont {Hase}, \citenamefont {Kay},\ and\
  \citenamefont {Ceotto}}]{Zhuang_Ceotto_Hessianapprox_2012}%
  \BibitemOpen
  \bibfield  {author} {\bibinfo {author} {\bibfnamefont {Y.}~\bibnamefont
  {Zhuang}}, \bibinfo {author} {\bibfnamefont {M.~R.}\ \bibnamefont {Siebert}},
  \bibinfo {author} {\bibfnamefont {W.~L.}\ \bibnamefont {Hase}}, \bibinfo
  {author} {\bibfnamefont {K.~G.}\ \bibnamefont {Kay}}, \ and\ \bibinfo
  {author} {\bibfnamefont {M.}~\bibnamefont {Ceotto}},\ }\href@noop {}
  {\bibfield  {journal} {\bibinfo  {journal} {J. Chem. Theory Comput.}\
  }\textbf {\bibinfo {volume} {9}},\ \bibinfo {pages} {54} (\bibinfo {year}
  {2012})}\BibitemShut {NoStop}%
\bibitem [{\citenamefont {Wehrle}\ \emph {et~al.}(2014)\citenamefont {Wehrle},
  \citenamefont {Sulc},\ and\ \citenamefont
  {Vanicek}}]{Wehrle_Vanicek_Oligothiophenes_2014}%
  \BibitemOpen
  \bibfield  {author} {\bibinfo {author} {\bibfnamefont {M.}~\bibnamefont
  {Wehrle}}, \bibinfo {author} {\bibfnamefont {M.}~\bibnamefont {Sulc}}, \ and\
  \bibinfo {author} {\bibfnamefont {J.}~\bibnamefont {Vanicek}},\ }\href
  {\doibase http://dx.doi.org/10.1063/1.4884718} {\bibfield  {journal}
  {\bibinfo  {journal} {J. Chem. Phys.}\ }\textbf {\bibinfo {volume} {140}},\
  \bibinfo {pages} {244114} (\bibinfo {year} {2014})}\BibitemShut {NoStop}%
\bibitem [{\citenamefont {Wehrle}\ \emph {et~al.}(2015)\citenamefont {Wehrle},
  \citenamefont {Oberli},\ and\ \citenamefont {Van{\'\i}{\v
  c}ek}}]{Wehrle_Vanicek_NH3_2015}%
  \BibitemOpen
  \bibfield  {author} {\bibinfo {author} {\bibfnamefont {M.}~\bibnamefont
  {Wehrle}}, \bibinfo {author} {\bibfnamefont {S.}~\bibnamefont {Oberli}}, \
  and\ \bibinfo {author} {\bibfnamefont {J.}~\bibnamefont {Van{\'\i}{\v
  c}ek}},\ }\href {\doibase 10.1021/acs.jpca.5b03907} {\bibfield  {journal}
  {\bibinfo  {journal} {J. Phys. Chem. A}\ }\textbf {\bibinfo {volume} {119}},\
  \bibinfo {pages} {5685} (\bibinfo {year} {2015})},\ \bibinfo {note} {pMID:
  25928833},\ \Eprint
  {http://arxiv.org/abs/http://dx.doi.org/10.1021/acs.jpca.5b03907}
  {http://dx.doi.org/10.1021/acs.jpca.5b03907} \BibitemShut {NoStop}%
\bibitem [{\citenamefont {Zambrano}\ \emph {et~al.}(2013)\citenamefont
  {Zambrano}, \citenamefont {{\v{S}}ulc},\ and\ \citenamefont
  {Van{\'\i}{\v{c}}ek}}]{Zambrano_Vanicek_Cellulardephasing_2013}%
  \BibitemOpen
  \bibfield  {author} {\bibinfo {author} {\bibfnamefont {E.}~\bibnamefont
  {Zambrano}}, \bibinfo {author} {\bibfnamefont {M.}~\bibnamefont
  {{\v{S}}ulc}}, \ and\ \bibinfo {author} {\bibfnamefont {J.}~\bibnamefont
  {Van{\'\i}{\v{c}}ek}},\ }\href@noop {} {\bibfield  {journal} {\bibinfo
  {journal} {J. Chem. Phys.}\ }\textbf {\bibinfo {volume} {139}},\ \bibinfo
  {pages} {054109} (\bibinfo {year} {2013})}\BibitemShut {NoStop}%
\bibitem [{\citenamefont {Braams}\ and\ \citenamefont
  {Bowman}(2009)}]{Braams_Bowman_PermutInvariant_2009}%
  \BibitemOpen
  \bibfield  {author} {\bibinfo {author} {\bibfnamefont {B.~J.}\ \bibnamefont
  {Braams}}\ and\ \bibinfo {author} {\bibfnamefont {J.~M.}\ \bibnamefont
  {Bowman}},\ }\href {\doibase 10.1080/01442350903234923} {\bibfield  {journal}
  {\bibinfo  {journal} {Int. Rev. Phys. Chem.}\ }\textbf {\bibinfo {volume}
  {28}},\ \bibinfo {pages} {577} (\bibinfo {year} {2009})}\BibitemShut
  {NoStop}%
\bibitem [{\citenamefont {Conte}\ \emph
  {et~al.}(2013{\natexlab{a}})\citenamefont {Conte}, \citenamefont {Fu},
  \citenamefont {Kamarchik},\ and\ \citenamefont
  {Bowman}}]{Conte_Bowman_GaussianBinning_2013}%
  \BibitemOpen
  \bibfield  {author} {\bibinfo {author} {\bibfnamefont {R.}~\bibnamefont
  {Conte}}, \bibinfo {author} {\bibfnamefont {B.}~\bibnamefont {Fu}}, \bibinfo
  {author} {\bibfnamefont {E.}~\bibnamefont {Kamarchik}}, \ and\ \bibinfo
  {author} {\bibfnamefont {J.~M.}\ \bibnamefont {Bowman}},\ }\href@noop {}
  {\bibfield  {journal} {\bibinfo  {journal} {J. Chem. Phys.}\ }\textbf
  {\bibinfo {volume} {139}},\ \bibinfo {pages} {044104} (\bibinfo {year}
  {2013}{\natexlab{a}})}\BibitemShut {NoStop}%
\bibitem [{\citenamefont {Jiang}\ and\ \citenamefont
  {Guo}(2014)}]{Jiang_Guo_NeuralNetworks_2014}%
  \BibitemOpen
  \bibfield  {author} {\bibinfo {author} {\bibfnamefont {B.}~\bibnamefont
  {Jiang}}\ and\ \bibinfo {author} {\bibfnamefont {H.}~\bibnamefont {Guo}},\
  }\href {\doibase http://dx.doi.org/10.1063/1.4887363} {\bibfield  {journal}
  {\bibinfo  {journal} {J. Chem. Phys.}\ }\textbf {\bibinfo {volume} {141}},\
  \bibinfo {pages} {034109} (\bibinfo {year} {2014})}\BibitemShut {NoStop}%
\bibitem [{\citenamefont {Conte}\ \emph {et~al.}(2014)\citenamefont {Conte},
  \citenamefont {Houston},\ and\ \citenamefont
  {Bowman}}]{Conte_Bowman_Communication_2014}%
  \BibitemOpen
  \bibfield  {author} {\bibinfo {author} {\bibfnamefont {R.}~\bibnamefont
  {Conte}}, \bibinfo {author} {\bibfnamefont {P.~L.}\ \bibnamefont {Houston}},
  \ and\ \bibinfo {author} {\bibfnamefont {J.~M.}\ \bibnamefont {Bowman}},\
  }\href@noop {} {\bibfield  {journal} {\bibinfo  {journal} {J. Chem. Phys.}\
  }\textbf {\bibinfo {volume} {140}},\ \bibinfo {pages} {151101} (\bibinfo
  {year} {2014})}\BibitemShut {NoStop}%
\bibitem [{\citenamefont {Houston}\ \emph {et~al.}(2014)\citenamefont
  {Houston}, \citenamefont {Conte},\ and\ \citenamefont
  {Bowman}}]{Houston_Bowman_CollisionModel_2014}%
  \BibitemOpen
  \bibfield  {author} {\bibinfo {author} {\bibfnamefont {P.~L.}\ \bibnamefont
  {Houston}}, \bibinfo {author} {\bibfnamefont {R.}~\bibnamefont {Conte}}, \
  and\ \bibinfo {author} {\bibfnamefont {J.~M.}\ \bibnamefont {Bowman}},\
  }\href@noop {} {\bibfield  {journal} {\bibinfo  {journal} {J. Phys. Chem. A}\
  }\textbf {\bibinfo {volume} {118}},\ \bibinfo {pages} {7758} (\bibinfo {year}
  {2014})}\BibitemShut {NoStop}%
\bibitem [{\citenamefont {Conte}\ \emph
  {et~al.}(2015{\natexlab{a}})\citenamefont {Conte}, \citenamefont {Houston},\
  and\ \citenamefont {Bowman}}]{Conte_Bowman_CollisionsCH4-H2O_2015}%
  \BibitemOpen
  \bibfield  {author} {\bibinfo {author} {\bibfnamefont {R.}~\bibnamefont
  {Conte}}, \bibinfo {author} {\bibfnamefont {P.~L.}\ \bibnamefont {Houston}},
  \ and\ \bibinfo {author} {\bibfnamefont {J.~M.}\ \bibnamefont {Bowman}},\
  }\href@noop {} {\bibfield  {journal} {\bibinfo  {journal} {J. Phys. Chem. A}\
  }\textbf {\bibinfo {volume} {119}},\ \bibinfo {pages} {12304} (\bibinfo
  {year} {2015}{\natexlab{a}})}\BibitemShut {NoStop}%
\bibitem [{\citenamefont {Conte}\ \emph
  {et~al.}(2015{\natexlab{b}})\citenamefont {Conte}, \citenamefont {Qu},\ and\
  \citenamefont {Bowman}}]{Conte_Bowman_Manybody_2015}%
  \BibitemOpen
  \bibfield  {author} {\bibinfo {author} {\bibfnamefont {R.}~\bibnamefont
  {Conte}}, \bibinfo {author} {\bibfnamefont {C.}~\bibnamefont {Qu}}, \ and\
  \bibinfo {author} {\bibfnamefont {J.~M.}\ \bibnamefont {Bowman}},\
  }\href@noop {} {\bibfield  {journal} {\bibinfo  {journal} {J. Chem. Theory
  Comp.}\ }\textbf {\bibinfo {volume} {11}},\ \bibinfo {pages} {1631} (\bibinfo
  {year} {2015}{\natexlab{b}})}\BibitemShut {NoStop}%
\bibitem [{\citenamefont {Houston}\ \emph {et~al.}(2015)\citenamefont
  {Houston}, \citenamefont {Conte},\ and\ \citenamefont
  {Bowman}}]{Houston_Bowman_ModelFinal_2015}%
  \BibitemOpen
  \bibfield  {author} {\bibinfo {author} {\bibfnamefont {P.~L.}\ \bibnamefont
  {Houston}}, \bibinfo {author} {\bibfnamefont {R.}~\bibnamefont {Conte}}, \
  and\ \bibinfo {author} {\bibfnamefont {J.~M.}\ \bibnamefont {Bowman}},\
  }\href@noop {} {\bibfield  {journal} {\bibinfo  {journal} {J. Phys. Chem. A}\
  }\textbf {\bibinfo {volume} {119}},\ \bibinfo {pages} {4695} (\bibinfo {year}
  {2015})}\BibitemShut {NoStop}%
\bibitem [{\citenamefont {Houston}\ \emph {et~al.}(2016)\citenamefont
  {Houston}, \citenamefont {Conte},\ and\ \citenamefont
  {Bowman}}]{Houston_Bowman_RoamingH2CO_2016}%
  \BibitemOpen
  \bibfield  {author} {\bibinfo {author} {\bibfnamefont {P.~L.}\ \bibnamefont
  {Houston}}, \bibinfo {author} {\bibfnamefont {R.}~\bibnamefont {Conte}}, \
  and\ \bibinfo {author} {\bibfnamefont {J.~M.}\ \bibnamefont {Bowman}},\
  }\href@noop {} {\bibfield  {journal} {\bibinfo  {journal} {J. Phys. Chem. A}\
  } (\bibinfo {year} {2016})}\BibitemShut {NoStop}%
\bibitem [{\citenamefont {Ceotto}\ \emph
  {et~al.}(2009{\natexlab{a}})\citenamefont {Ceotto}, \citenamefont {Atahan},
  \citenamefont {Tantardini},\ and\ \citenamefont
  {Aspuru-Guzik}}]{Ceotto_AspuruGuzik_Multiplecoherent_2009}%
  \BibitemOpen
  \bibfield  {author} {\bibinfo {author} {\bibfnamefont {M.}~\bibnamefont
  {Ceotto}}, \bibinfo {author} {\bibfnamefont {S.}~\bibnamefont {Atahan}},
  \bibinfo {author} {\bibfnamefont {G.~F.}\ \bibnamefont {Tantardini}}, \ and\
  \bibinfo {author} {\bibfnamefont {A.}~\bibnamefont {Aspuru-Guzik}},\ }\href
  {\doibase http://dx.doi.org/10.1063/1.3155062} {\bibfield  {journal}
  {\bibinfo  {journal} {J. Chem. Phys.}\ }\textbf {\bibinfo {volume} {130}},\
  \bibinfo {pages} {234113} (\bibinfo {year} {2009}{\natexlab{a}})}\BibitemShut
  {NoStop}%
\bibitem [{\citenamefont {Tatchen}\ and\ \citenamefont
  {Pollak}(2009)}]{Tatchen_Pollak_Onthefly_2009}%
  \BibitemOpen
  \bibfield  {author} {\bibinfo {author} {\bibfnamefont {J.}~\bibnamefont
  {Tatchen}}\ and\ \bibinfo {author} {\bibfnamefont {E.}~\bibnamefont
  {Pollak}},\ }\href {\doibase http://dx.doi.org/10.1063/1.3074100} {\bibfield
  {journal} {\bibinfo  {journal} {J. Chem. Phys.}\ }\textbf {\bibinfo {volume}
  {130}},\ \bibinfo {pages} {041103} (\bibinfo {year} {2009})}\BibitemShut
  {NoStop}%
\bibitem [{\citenamefont {Ceotto}\ \emph
  {et~al.}(2011{\natexlab{a}})\citenamefont {Ceotto}, \citenamefont
  {Tantardini},\ and\ \citenamefont
  {Aspuru-Guzik}}]{Ceotto_AspuruGuzik_Curseofdimensionality_2011}%
  \BibitemOpen
  \bibfield  {author} {\bibinfo {author} {\bibfnamefont {M.}~\bibnamefont
  {Ceotto}}, \bibinfo {author} {\bibfnamefont {G.~F.}\ \bibnamefont
  {Tantardini}}, \ and\ \bibinfo {author} {\bibfnamefont {A.}~\bibnamefont
  {Aspuru-Guzik}},\ }\href {\doibase http://dx.doi.org/10.1063/1.3664731}
  {\bibfield  {journal} {\bibinfo  {journal} {J. Chem. Phys.}\ }\textbf
  {\bibinfo {volume} {135}},\ \bibinfo {pages} {214108} (\bibinfo {year}
  {2011}{\natexlab{a}})}\BibitemShut {NoStop}%
\bibitem [{\citenamefont {Wong}\ \emph {et~al.}(2011)\citenamefont {Wong},
  \citenamefont {Benoit}, \citenamefont {Lewerenz}, \citenamefont {Brown},\
  and\ \citenamefont {Roy}}]{Wong_Roy_Formaldehyde_2011}%
  \BibitemOpen
  \bibfield  {author} {\bibinfo {author} {\bibfnamefont {S.~Y.~Y.}\
  \bibnamefont {Wong}}, \bibinfo {author} {\bibfnamefont {D.~M.}\ \bibnamefont
  {Benoit}}, \bibinfo {author} {\bibfnamefont {M.}~\bibnamefont {Lewerenz}},
  \bibinfo {author} {\bibfnamefont {A.}~\bibnamefont {Brown}}, \ and\ \bibinfo
  {author} {\bibfnamefont {P.-N.}\ \bibnamefont {Roy}},\ }\href {\doibase
  http://dx.doi.org/10.1063/1.3553179} {\bibfield  {journal} {\bibinfo
  {journal} {J. Chem. Phys.}\ }\textbf {\bibinfo {volume} {134}},\ \bibinfo
  {pages} {094110} (\bibinfo {year} {2011})}\BibitemShut {NoStop}%
\bibitem [{\citenamefont {Ceotto}\ \emph {et~al.}(2013)\citenamefont {Ceotto},
  \citenamefont {Zhuang},\ and\ \citenamefont
  {Hase}}]{Ceotto_Hase_AcceleratedSC_2013}%
  \BibitemOpen
  \bibfield  {author} {\bibinfo {author} {\bibfnamefont {M.}~\bibnamefont
  {Ceotto}}, \bibinfo {author} {\bibfnamefont {Y.}~\bibnamefont {Zhuang}}, \
  and\ \bibinfo {author} {\bibfnamefont {W.~L.}\ \bibnamefont {Hase}},\
  }\href@noop {} {\bibfield  {journal} {\bibinfo  {journal} {J. Chem. Phys.}\
  }\textbf {\bibinfo {volume} {138}},\ \bibinfo {pages} {054116} (\bibinfo
  {year} {2013})}\BibitemShut {NoStop}%
\bibitem [{\citenamefont {Heller}(1981)}]{Heller_FrozenGaussian_1981}%
  \BibitemOpen
  \bibfield  {author} {\bibinfo {author} {\bibfnamefont {E.~J.}\ \bibnamefont
  {Heller}},\ }\href {\doibase http://dx.doi.org/10.1063/1.442382} {\bibfield
  {journal} {\bibinfo  {journal} {J. Chem. Phys.}\ }\textbf {\bibinfo {volume}
  {75}},\ \bibinfo {pages} {2923} (\bibinfo {year} {1981})}\BibitemShut
  {NoStop}%
\bibitem [{\citenamefont {Herman}\ and\ \citenamefont
  {Kluk}(1984)}]{Herman_Kluk_SCnonspreading_1984}%
  \BibitemOpen
  \bibfield  {author} {\bibinfo {author} {\bibfnamefont {M.~F.}\ \bibnamefont
  {Herman}}\ and\ \bibinfo {author} {\bibfnamefont {E.}~\bibnamefont {Kluk}},\
  }\href {\doibase http://dx.doi.org/10.1016/0301-0104(84)80039-7} {\bibfield
  {journal} {\bibinfo  {journal} {Chem. Phys.}\ }\textbf {\bibinfo {volume}
  {91}},\ \bibinfo {pages} {27 } (\bibinfo {year} {1984})}\BibitemShut
  {NoStop}%
\bibitem [{\citenamefont {Miller}(2005)}]{Miller_PNAScomplexsystems_2005}%
  \BibitemOpen
  \bibfield  {author} {\bibinfo {author} {\bibfnamefont {W.~H.}\ \bibnamefont
  {Miller}},\ }\href {\doibase 10.1073/pnas.0408043102} {\bibfield  {journal}
  {\bibinfo  {journal} {Proc. Natl. Acad. Sci. USA}\ }\textbf {\bibinfo
  {volume} {102}},\ \bibinfo {pages} {6660} (\bibinfo {year}
  {2005})}\BibitemShut {NoStop}%
\bibitem [{\citenamefont {Kay}(2005)}]{Kay_Atomsandmolecules_2005}%
  \BibitemOpen
  \bibfield  {author} {\bibinfo {author} {\bibfnamefont {K.~G.}\ \bibnamefont
  {Kay}},\ }\href@noop {} {\bibfield  {journal} {\bibinfo  {journal} {Annu.
  Rev. Phys. Chem.}\ }\textbf {\bibinfo {volume} {56}},\ \bibinfo {pages} {255}
  (\bibinfo {year} {2005})}\BibitemShut {NoStop}%
\bibitem [{\citenamefont
  {Miller}(1970{\natexlab{a}})}]{Miller_Atom-Diatom_1970}%
  \BibitemOpen
  \bibfield  {author} {\bibinfo {author} {\bibfnamefont {W.~H.}\ \bibnamefont
  {Miller}},\ }\href {\doibase http://dx.doi.org/10.1063/1.1674275} {\bibfield
  {journal} {\bibinfo  {journal} {J. Chem. Phys.}\ }\textbf {\bibinfo {volume}
  {53}},\ \bibinfo {pages} {1949} (\bibinfo {year}
  {1970}{\natexlab{a}})}\BibitemShut {NoStop}%
\bibitem [{\citenamefont {Miller}(1974)}]{Miller_Molecularcollisions_1974}%
  \BibitemOpen
  \bibfield  {author} {\bibinfo {author} {\bibfnamefont {W.~H.}\ \bibnamefont
  {Miller}},\ }\href@noop {} {\bibfield  {journal} {\bibinfo  {journal} {Adv.
  Chem. Phys}\ }\textbf {\bibinfo {volume} {25}},\ \bibinfo {pages} {69}
  (\bibinfo {year} {1974})}\BibitemShut {NoStop}%
\bibitem [{\citenamefont {Ceotto}\ \emph {et~al.}(2010)\citenamefont {Ceotto},
  \citenamefont {Dell`~Angelo},\ and\ \citenamefont
  {Tantardini}}]{Ceotto_Tantardini_Copper100_2010}%
  \BibitemOpen
  \bibfield  {author} {\bibinfo {author} {\bibfnamefont {M.}~\bibnamefont
  {Ceotto}}, \bibinfo {author} {\bibfnamefont {D.}~\bibnamefont
  {Dell`~Angelo}}, \ and\ \bibinfo {author} {\bibfnamefont {G.~F.}\
  \bibnamefont {Tantardini}},\ }\href@noop {} {\bibfield  {journal} {\bibinfo
  {journal} {J. Chem. Phys.}\ }\textbf {\bibinfo {volume} {133}},\ \bibinfo
  {pages} {054701} (\bibinfo {year} {2010})}\BibitemShut {NoStop}%
\bibitem [{\citenamefont {De~Leon}\ and\ \citenamefont
  {Heller}(1983)}]{DeLeon_Heller_SCeigenfunctions_1983}%
  \BibitemOpen
  \bibfield  {author} {\bibinfo {author} {\bibfnamefont {N.}~\bibnamefont
  {De~Leon}}\ and\ \bibinfo {author} {\bibfnamefont {E.~J.}\ \bibnamefont
  {Heller}},\ }\href@noop {} {\bibfield  {journal} {\bibinfo  {journal} {J.
  Chem. Phys.}\ }\textbf {\bibinfo {volume} {78}},\ \bibinfo {pages} {4005}
  (\bibinfo {year} {1983})}\BibitemShut {NoStop}%
\bibitem [{\citenamefont {Ceotto}\ \emph
  {et~al.}(2009{\natexlab{b}})\citenamefont {Ceotto}, \citenamefont {Atahan},
  \citenamefont {Shim}, \citenamefont {Tantardini},\ and\ \citenamefont
  {Aspuru-Guzik}}]{Ceotto_AspuruGuzik_PCCPFirstprinciples_2009}%
  \BibitemOpen
  \bibfield  {author} {\bibinfo {author} {\bibfnamefont {M.}~\bibnamefont
  {Ceotto}}, \bibinfo {author} {\bibfnamefont {S.}~\bibnamefont {Atahan}},
  \bibinfo {author} {\bibfnamefont {S.}~\bibnamefont {Shim}}, \bibinfo {author}
  {\bibfnamefont {G.~F.}\ \bibnamefont {Tantardini}}, \ and\ \bibinfo {author}
  {\bibfnamefont {A.}~\bibnamefont {Aspuru-Guzik}},\ }\href {\doibase
  10.1039/B820785B} {\bibfield  {journal} {\bibinfo  {journal} {Phys. Chem.
  Chem. Phys.}\ }\textbf {\bibinfo {volume} {11}},\ \bibinfo {pages} {3861}
  (\bibinfo {year} {2009}{\natexlab{b}})}\BibitemShut {NoStop}%
\bibitem [{\citenamefont {Ceotto}\ \emph
  {et~al.}(2011{\natexlab{b}})\citenamefont {Ceotto}, \citenamefont {Valleau},
  \citenamefont {Tantardini},\ and\ \citenamefont
  {Aspuru-Guzik}}]{Ceotto_AspuruGuzik_Firstprinciples_2011}%
  \BibitemOpen
  \bibfield  {author} {\bibinfo {author} {\bibfnamefont {M.}~\bibnamefont
  {Ceotto}}, \bibinfo {author} {\bibfnamefont {S.}~\bibnamefont {Valleau}},
  \bibinfo {author} {\bibfnamefont {G.~F.}\ \bibnamefont {Tantardini}}, \ and\
  \bibinfo {author} {\bibfnamefont {A.}~\bibnamefont {Aspuru-Guzik}},\ }\href
  {\doibase http://dx.doi.org/10.1063/1.3599469} {\bibfield  {journal}
  {\bibinfo  {journal} {J. Chem. Phys.}\ }\textbf {\bibinfo {volume} {134}},\
  \bibinfo {pages} {234103} (\bibinfo {year} {2011}{\natexlab{b}})}\BibitemShut
  {NoStop}%
\bibitem [{\citenamefont {Conte}\ \emph
  {et~al.}(2013{\natexlab{b}})\citenamefont {Conte}, \citenamefont
  {Aspuru-Guzik},\ and\ \citenamefont {Ceotto}}]{Conte_Ceotto_NH3_2013}%
  \BibitemOpen
  \bibfield  {author} {\bibinfo {author} {\bibfnamefont {R.}~\bibnamefont
  {Conte}}, \bibinfo {author} {\bibfnamefont {A.}~\bibnamefont {Aspuru-Guzik}},
  \ and\ \bibinfo {author} {\bibfnamefont {M.}~\bibnamefont {Ceotto}},\ }\href
  {\doibase 10.1021/jz401603f} {\bibfield  {journal} {\bibinfo  {journal} {J.
  Phys. Chem. Lett.}\ }\textbf {\bibinfo {volume} {4}},\ \bibinfo {pages}
  {3407} (\bibinfo {year} {2013}{\natexlab{b}})}\BibitemShut {NoStop}%
\bibitem [{\citenamefont {Tamascelli}\ \emph {et~al.}(2014)\citenamefont
  {Tamascelli}, \citenamefont {Dambrosio}, \citenamefont {Conte},\ and\
  \citenamefont {Ceotto}}]{Tamascelli_Ceotto_GPU_2014}%
  \BibitemOpen
  \bibfield  {author} {\bibinfo {author} {\bibfnamefont {D.}~\bibnamefont
  {Tamascelli}}, \bibinfo {author} {\bibfnamefont {F.~S.}\ \bibnamefont
  {Dambrosio}}, \bibinfo {author} {\bibfnamefont {R.}~\bibnamefont {Conte}}, \
  and\ \bibinfo {author} {\bibfnamefont {M.}~\bibnamefont {Ceotto}},\
  }\href@noop {} {\bibfield  {journal} {\bibinfo  {journal} {J. Chem. Phys.}\
  }\textbf {\bibinfo {volume} {140}},\ \bibinfo {pages} {174109} (\bibinfo
  {year} {2014})}\BibitemShut {NoStop}%
\bibitem [{\citenamefont {Buchholz}\ \emph {et~al.}(2016)\citenamefont
  {Buchholz}, \citenamefont {Grossmann},\ and\ \citenamefont
  {Ceotto}}]{Buchholz_Ceotto_MixedSC_2016}%
  \BibitemOpen
  \bibfield  {author} {\bibinfo {author} {\bibfnamefont {M.}~\bibnamefont
  {Buchholz}}, \bibinfo {author} {\bibfnamefont {F.}~\bibnamefont {Grossmann}},
  \ and\ \bibinfo {author} {\bibfnamefont {M.}~\bibnamefont {Ceotto}},\
  }\href@noop {} {\bibfield  {journal} {\bibinfo  {journal} {J. Chem. Phys.}\
  }\textbf {\bibinfo {volume} {144}},\ \bibinfo {pages} {094102} (\bibinfo
  {year} {2016})}\BibitemShut {NoStop}%
\bibitem [{\citenamefont {Di~Liberto}\ and\ \citenamefont
  {Ceotto}(2016)}]{DiLiberto_Ceotto_Prefactors_2016}%
  \BibitemOpen
  \bibfield  {author} {\bibinfo {author} {\bibfnamefont {G.}~\bibnamefont
  {Di~Liberto}}\ and\ \bibinfo {author} {\bibfnamefont {M.}~\bibnamefont
  {Ceotto}},\ }\href@noop {} {\bibfield  {journal} {\bibinfo  {journal} {J.
  Chem. Phys.}\ }\textbf {\bibinfo {volume} {145}},\ \bibinfo {pages} {144107}
  (\bibinfo {year} {2016})}\BibitemShut {NoStop}%
\bibitem [{\citenamefont {Trumpler}\ and\ \citenamefont
  {Weaver}(1962)}]{Trumpler_Weaver_StatAstronomy_1962}%
  \BibitemOpen
  \bibfield  {author} {\bibinfo {author} {\bibfnamefont {R.~J.}\ \bibnamefont
  {Trumpler}}\ and\ \bibinfo {author} {\bibfnamefont {H.~F.}\ \bibnamefont
  {Weaver}},\ }\href@noop {} {\emph {\bibinfo {title} {Statistical
  astronomy}}}\ (\bibinfo  {publisher} {Dover Publications},\ \bibinfo {year}
  {1962})\BibitemShut {NoStop}%
\bibitem [{\citenamefont {Gutzwiller}(1967)}]{Gutzwiller_SCpropagator_1967}%
  \BibitemOpen
  \bibfield  {author} {\bibinfo {author} {\bibfnamefont {M.~C.}\ \bibnamefont
  {Gutzwiller}},\ }\href@noop {} {\bibfield  {journal} {\bibinfo  {journal} {J.
  Math. Phys.}\ }\textbf {\bibinfo {volume} {8}},\ \bibinfo {pages} {1979}
  (\bibinfo {year} {1967})}\BibitemShut {NoStop}%
\bibitem [{\citenamefont {Van~Vleck}(1928)}]{VanVleck_SCpropagator_1928}%
  \BibitemOpen
  \bibfield  {author} {\bibinfo {author} {\bibfnamefont {J.~H.}\ \bibnamefont
  {Van~Vleck}},\ }\href@noop {} {\bibfield  {journal} {\bibinfo  {journal}
  {Proc. Natl. Acad. Sci.}\ }\textbf {\bibinfo {volume} {14}},\ \bibinfo
  {pages} {178} (\bibinfo {year} {1928})}\BibitemShut {NoStop}%
\bibitem [{\citenamefont {Berry}\ and\ \citenamefont
  {Mount}(1972)}]{Berry_Mount_Semiclassical_1972}%
  \BibitemOpen
  \bibfield  {author} {\bibinfo {author} {\bibfnamefont {M.~V.}\ \bibnamefont
  {Berry}}\ and\ \bibinfo {author} {\bibfnamefont {K.}~\bibnamefont {Mount}},\
  }\href@noop {} {\bibfield  {journal} {\bibinfo  {journal} {Rep. on Prog.
  Phys.}\ }\textbf {\bibinfo {volume} {35}},\ \bibinfo {pages} {315} (\bibinfo
  {year} {1972})}\BibitemShut {NoStop}%
\bibitem [{\citenamefont {Miller}(1970{\natexlab{b}})}]{Miller_S-Matrix_1970}%
  \BibitemOpen
  \bibfield  {author} {\bibinfo {author} {\bibfnamefont {W.~H.}\ \bibnamefont
  {Miller}},\ }\href {\doibase http://dx.doi.org/10.1063/1.1674535} {\bibfield
  {journal} {\bibinfo  {journal} {J. Chem. Phys.}\ }\textbf {\bibinfo {volume}
  {53}},\ \bibinfo {pages} {3578} (\bibinfo {year}
  {1970}{\natexlab{b}})}\BibitemShut {NoStop}%
\bibitem [{\citenamefont {Heller}(1991)}]{Heller_CellularDynamics_1991}%
  \BibitemOpen
  \bibfield  {author} {\bibinfo {author} {\bibfnamefont {E.~J.}\ \bibnamefont
  {Heller}},\ }\href {\doibase http://dx.doi.org/10.1063/1.459848} {\bibfield
  {journal} {\bibinfo  {journal} {J. Chem. Phys.}\ }\textbf {\bibinfo {volume}
  {94}},\ \bibinfo {pages} {2723} (\bibinfo {year} {1991})}\BibitemShut
  {NoStop}%
\bibitem [{\citenamefont {Weissman}(1982)}]{Weissman_coherentstates_1982}%
  \BibitemOpen
  \bibfield  {author} {\bibinfo {author} {\bibfnamefont {Y.}~\bibnamefont
  {Weissman}},\ }\href@noop {} {\bibfield  {journal} {\bibinfo  {journal} {J.
  Chem. Phys.}\ }\textbf {\bibinfo {volume} {76}},\ \bibinfo {pages} {4067}
  (\bibinfo {year} {1982})}\BibitemShut {NoStop}%
\bibitem [{\citenamefont {Baranger}\ \emph {et~al.}(2001)\citenamefont
  {Baranger}, \citenamefont {de~Aguiar}, \citenamefont {Keck}, \citenamefont
  {Korsch},\ and\ \citenamefont {Schellhaass}}]{Baranger_Schellhaass_2001}%
  \BibitemOpen
  \bibfield  {author} {\bibinfo {author} {\bibfnamefont {M.}~\bibnamefont
  {Baranger}}, \bibinfo {author} {\bibfnamefont {M.~A.}\ \bibnamefont
  {de~Aguiar}}, \bibinfo {author} {\bibfnamefont {F.}~\bibnamefont {Keck}},
  \bibinfo {author} {\bibfnamefont {H.-J.}\ \bibnamefont {Korsch}}, \ and\
  \bibinfo {author} {\bibfnamefont {B.}~\bibnamefont {Schellhaass}},\
  }\href@noop {} {\bibfield  {journal} {\bibinfo  {journal} {J. Phys. A}\
  }\textbf {\bibinfo {volume} {34}},\ \bibinfo {pages} {7227} (\bibinfo {year}
  {2001})}\BibitemShut {NoStop}%
\bibitem [{\citenamefont {Kaledin}\ and\ \citenamefont
  {Miller}(2003{\natexlab{a}})}]{Kaledin_Miller_TAmolecules_2003}%
  \BibitemOpen
  \bibfield  {author} {\bibinfo {author} {\bibfnamefont {A.~L.}\ \bibnamefont
  {Kaledin}}\ and\ \bibinfo {author} {\bibfnamefont {W.~H.}\ \bibnamefont
  {Miller}},\ }\href {\doibase http://dx.doi.org/10.1063/1.1589477} {\bibfield
  {journal} {\bibinfo  {journal} {J. Chem. Phys.}\ }\textbf {\bibinfo {volume}
  {119}},\ \bibinfo {pages} {3078} (\bibinfo {year}
  {2003}{\natexlab{a}})}\BibitemShut {NoStop}%
\bibitem [{\citenamefont {Kaledin}\ and\ \citenamefont
  {Miller}(2003{\natexlab{b}})}]{Kaledin_Miller_Timeaveraging_2003}%
  \BibitemOpen
  \bibfield  {author} {\bibinfo {author} {\bibfnamefont {A.~L.}\ \bibnamefont
  {Kaledin}}\ and\ \bibinfo {author} {\bibfnamefont {W.~H.}\ \bibnamefont
  {Miller}},\ }\href {\doibase http://dx.doi.org/10.1063/1.1562158} {\bibfield
  {journal} {\bibinfo  {journal} {J. Chem. Phys.}\ }\textbf {\bibinfo {volume}
  {118}},\ \bibinfo {pages} {7174} (\bibinfo {year}
  {2003}{\natexlab{b}})}\BibitemShut {NoStop}%
\bibitem [{\citenamefont {Gabas}\ \emph {et~al.}(2017)\citenamefont {Gabas},
  \citenamefont {Conte},\ and\ \citenamefont
  {Ceotto}}]{Gabas_Ceotto_Glycine_2016}%
  \BibitemOpen
  \bibfield  {author} {\bibinfo {author} {\bibfnamefont {F.}~\bibnamefont
  {Gabas}}, \bibinfo {author} {\bibfnamefont {R.}~\bibnamefont {Conte}}, \ and\
  \bibinfo {author} {\bibfnamefont {M.}~\bibnamefont {Ceotto}},\ }\href@noop {}
  {\bibfield  {journal} {\bibinfo  {journal} {J. Chem. Theory Comp.}\ }\textbf
  {\bibinfo {volume} {under review}} (\bibinfo {year} {2017})}\BibitemShut
  {NoStop}%
\bibitem [{\citenamefont {Tannor}(2007)}]{Tannor_book_2007}%
  \BibitemOpen
  \bibfield  {author} {\bibinfo {author} {\bibfnamefont {D.~J.}\ \bibnamefont
  {Tannor}},\ }\href@noop {} {\emph {\bibinfo {title} {Introduction to quantum
  mechanics}}}\ (\bibinfo  {publisher} {University Science Books},\ \bibinfo
  {year} {2007})\BibitemShut {NoStop}%
\bibitem [{Sup()}]{Suppl_Info_PRL}%
  \BibitemOpen
  \href@noop {} {\ }\bibinfo {note} {See Supplemental Material at [] for a
  detailed derivation of formulae (1)-(6).}\BibitemShut {Stop}%
\bibitem [{\citenamefont {Hinsen}\ and\ \citenamefont
  {Kneller}(2000)}]{Hinsen_Kneller_SingValueDecomp_2000}%
  \BibitemOpen
  \bibfield  {author} {\bibinfo {author} {\bibfnamefont {K.}~\bibnamefont
  {Hinsen}}\ and\ \bibinfo {author} {\bibfnamefont {G.~R.}\ \bibnamefont
  {Kneller}},\ }\href {\doibase 10.1080/08927020008025373} {\bibfield
  {journal} {\bibinfo  {journal} {Mol. Simul.}\ }\textbf {\bibinfo {volume}
  {23}},\ \bibinfo {pages} {275} (\bibinfo {year} {2000})}\BibitemShut
  {NoStop}%
\bibitem [{\citenamefont {Harland}\ and\ \citenamefont
  {Roy}(2003)}]{Harland_Roy_SCIVRconstrained_2003}%
  \BibitemOpen
  \bibfield  {author} {\bibinfo {author} {\bibfnamefont {B.~B.}\ \bibnamefont
  {Harland}}\ and\ \bibinfo {author} {\bibfnamefont {P.-N.}\ \bibnamefont
  {Roy}},\ }\href {\doibase http://dx.doi.org/10.1063/1.1545772} {\bibfield
  {journal} {\bibinfo  {journal} {J. Chem. Phys.}\ }\textbf {\bibinfo {volume}
  {118}},\ \bibinfo {pages} {4791} (\bibinfo {year} {2003})}\BibitemShut
  {NoStop}%
\bibitem [{\citenamefont {Bowman}\ \emph {et~al.}(1988)\citenamefont {Bowman},
  \citenamefont {Wierzbicki},\ and\ \citenamefont
  {Zuniga}}]{Bowman_Zuniga_H2Opotential_1988}%
  \BibitemOpen
  \bibfield  {author} {\bibinfo {author} {\bibfnamefont {J.~M.}\ \bibnamefont
  {Bowman}}, \bibinfo {author} {\bibfnamefont {A.}~\bibnamefont {Wierzbicki}},
  \ and\ \bibinfo {author} {\bibfnamefont {J.}~\bibnamefont {Zuniga}},\ }\href
  {\doibase http://dx.doi.org/10.1016/0009-2614(88)80040-X} {\bibfield
  {journal} {\bibinfo  {journal} {Chem. Phys. Lett.}\ }\textbf {\bibinfo
  {volume} {150}},\ \bibinfo {pages} {269 } (\bibinfo {year}
  {1988})}\BibitemShut {NoStop}%
\bibitem [{\citenamefont {Carter}\ \emph {et~al.}(1999)\citenamefont {Carter},
  \citenamefont {Shnider},\ and\ \citenamefont
  {Bowman}}]{Carter_Bowman_Methane_1999}%
  \BibitemOpen
  \bibfield  {author} {\bibinfo {author} {\bibfnamefont {S.}~\bibnamefont
  {Carter}}, \bibinfo {author} {\bibfnamefont {H.~M.}\ \bibnamefont {Shnider}},
  \ and\ \bibinfo {author} {\bibfnamefont {J.~M.}\ \bibnamefont {Bowman}},\
  }\href@noop {} {\bibfield  {journal} {\bibinfo  {journal} {J. Chem. Phys.}\
  }\textbf {\bibinfo {volume} {110}},\ \bibinfo {pages} {8417} (\bibinfo {year}
  {1999})}\BibitemShut {NoStop}%
\bibitem [{\citenamefont {Maslen}\ \emph {et~al.}(1992)\citenamefont {Maslen},
  \citenamefont {Handy}, \citenamefont {Amos},\ and\ \citenamefont
  {Jayatilaka}}]{Maslen_Jayatilaka_Anharmonicbenzene_1992}%
  \BibitemOpen
  \bibfield  {author} {\bibinfo {author} {\bibfnamefont {P.~E.}\ \bibnamefont
  {Maslen}}, \bibinfo {author} {\bibfnamefont {N.~C.}\ \bibnamefont {Handy}},
  \bibinfo {author} {\bibfnamefont {R.~D.}\ \bibnamefont {Amos}}, \ and\
  \bibinfo {author} {\bibfnamefont {D.}~\bibnamefont {Jayatilaka}},\ }\href
  {\doibase http://dx.doi.org/10.1063/1.463926} {\bibfield  {journal} {\bibinfo
   {journal} {J. Chem. Phys.}\ }\textbf {\bibinfo {volume} {97}},\ \bibinfo
  {pages} {4233} (\bibinfo {year} {1992})}\BibitemShut {NoStop}%
\bibitem [{\citenamefont {Halverson}\ and\ \citenamefont
  {Poirier}(2015)}]{Halverson_Poirier_Benzene_2015}%
  \BibitemOpen
  \bibfield  {author} {\bibinfo {author} {\bibfnamefont {T.}~\bibnamefont
  {Halverson}}\ and\ \bibinfo {author} {\bibfnamefont {B.}~\bibnamefont
  {Poirier}},\ }\href@noop {} {\bibfield  {journal} {\bibinfo  {journal} {J.
  Phys. Chem. A}\ }\textbf {\bibinfo {volume} {119}},\ \bibinfo {pages} {12417}
  (\bibinfo {year} {2015})}\BibitemShut {NoStop}%
\bibitem [{\citenamefont {Holec}\ \emph {et~al.}(2010)\citenamefont {Holec},
  \citenamefont {Hartmann}, \citenamefont {Fischer}, \citenamefont
  {Rammerstorfer}, \citenamefont {Mayrhofer},\ and\ \citenamefont
  {Paris}}]{Holec_Paris_FullereneFF_2010}%
  \BibitemOpen
  \bibfield  {author} {\bibinfo {author} {\bibfnamefont {D.}~\bibnamefont
  {Holec}}, \bibinfo {author} {\bibfnamefont {M.~A.}\ \bibnamefont {Hartmann}},
  \bibinfo {author} {\bibfnamefont {F.~D.}\ \bibnamefont {Fischer}}, \bibinfo
  {author} {\bibfnamefont {F.~G.}\ \bibnamefont {Rammerstorfer}}, \bibinfo
  {author} {\bibfnamefont {P.~H.}\ \bibnamefont {Mayrhofer}}, \ and\ \bibinfo
  {author} {\bibfnamefont {O.}~\bibnamefont {Paris}},\ }\href {\doibase
  10.1103/PhysRevB.81.235403} {\bibfield  {journal} {\bibinfo  {journal} {Phys.
  Rev. B}\ }\textbf {\bibinfo {volume} {81}},\ \bibinfo {pages} {235403}
  (\bibinfo {year} {2010})}\BibitemShut {NoStop}%
\end{thebibliography}%

\end{document}